\documentclass[default, astrosymb, twocolumn]{aastex701}

\pdfoutput=1
\usepackage{graphicx} % Required for inserting images
\usepackage{natbib}
\usepackage{float}
\usepackage[caption = false]{subfig}

\newcommand{\msol}{M$_\odot\ $}

\begin{document}

\title{Studying the Variable Continuum and Emission Line Spectrum of 4U 1700-377 using NuSTAR’s Stray Light}

\correspondingauthor{C. J. Bourque}

\author[orcid=0009-0002-8981-2197,sname='Bourque', gname='Colin']{C. J. Bourque}
\affiliation{Department of Physics and Astronomy, Middlebury College, Middlebury, VT 05753, USA}
\email{cbourque@wayne.edu}

\author[orcid=0000-0002-4024-6967,sname='Brumback', gname='McKinley']{M. C. Brumback}
\affiliation{Department of Physics and Astronomy, Middlebury College, Middlebury, VT 05753, USA}
\email{mbrumback@middlebury.edu}

\author[orcid=0000-0002-1984-2932,sname='Grefenstette', gname='Brian']{B. W. Grefenstette}
\affiliation{Cahill Center for Astronomy and Astrophysics, California Institute of Technology, Pasadena, CA 91125, USA}
\email{bwgref@srl.caltech.edu}

\author[orcid=0000-0002-5341-6929,sname='Buisson', gname='Brian']{D. J. K. Buisson}
\affiliation{Independent Researcher}
\email{djkbuisson@gmail.com}

\author[orcid=0000-0002-8961-939X,sname='Ludlam', gname='Renee']{R. M. Ludlam}
\affiliation{Department of Physics \& Astronomy, Wayne State University, 666 West Hancock Street, Detroit, MI 48201, USA}
\email{ef2051@wayne.edu}

\author[orcid=0000-0003-4216-7936,sname='Mastroserio', gname='Guglielmo']{G. Mastroserio}
\affiliation{Scuola Universitaria Superiore IUSS Pavia, Palazzo del Broletto, piazza della Vittoria 15, I-27100 Pavia, Italy}
\email{guglielmo.mastroserio@iusspavia.it}

\author[orcid=0000-0001-6304-1035,sname='Rossland', gname='Steven']{S. Rossland}
\affiliation{Center for National Security Initiatives, University of Colorado Boulder, Boulder, CO 80309, USA}
\email{coimhead1643@gmail.com}

\email[show]{cbourque@wayne.edu}  

\begin{abstract}

The high-mass X-ray binary 4U 1700-377 shows strong variation in brightness on timescales of hundreds of seconds due to accretion from a highly clumped stellar wind from the companion. Using two focused observations and five stray light observations from NuSTAR, we are able to expand the baseline of observations of the source and track spectral parameters throughout the source's variability. The focused NuSTAR observations confidently detect excess emission lines at energies above Fe K$\alpha$, but are unable to universally constrain the nature of these lines independently of model choice. Strong Fe K$\alpha$ fluorescence is ubiquitous in the stray light data, but these observations lack sufficient signal-to-noise to further measure additional emission lines. We discuss the statistics of a claimed cyclotron resonant scattering feature as seen in the stray light spectra, but find no conclusive evidence for its existence. Finally, we investigate two instances of sustained low-flux states of up to 60 ks. When this occurs, luminosity can fall by almost an order of magnitude and the high-energy continuum softens significantly without corresponding to increased absorption. We discuss the manner in which the changing shape of the continuum during these intervals may show accretion during an extended rarefied interval of stellar wind, or a possible change of the dominant accretion regime.

\end{abstract}

%%%%%%%%%%%%%%%%%%%%%%%%%%%%%%%%%%%%%%%%%%%%%%%%%%%%%%%%%%%%%%%%%%%%%%%%%%%
%%%%%%%%%%%%%%%%%%%%%%%%%%%%%%%%%%%%%%%%%%%%%%%%%%%%%%%%%%%%%%%%%%%%%%%%%%%
\section{Introduction}

The high mass X-ray binary (HMXB) \object{4U 1700-377} \citep{Jones1973} is an eclipsing binary system containing a massive (type O6f) companion star \citep{Hainich2020} and a compact object, the nature of which is uncertain. The orbital period is reported to be 3.41 days \citep{IslamPaul2016}. The source does not appear to exhibit pulsations \citep{Dolan1980, Xiao2024, WO2026}, and the presence of a cyclotron resonant scattering feature (CRSF) has been debated \citep[see][]{Reynolds1999, Jiaswal2015a, Seifina2016, Bala2020, Xiao2024, WO2026}. In spite of this, it is generally believed that the compact object in 4U 1700-377 is a neutron star, due to spectral indications of Compton cooling \citep{Chicharro2018} and due to quasi-stability of the photon index during hard-soft state transitions \citep{Seifina2016}. 

Accretion onto the compact object in 4U 1700-377 is driven by the stellar wind from its massive companion \citep{Hainich2020}. Since stellar wind outflow is unstable with respect to velocity perturbations, this leads to accretion in the form of dense ``clumps" \citep{MN2017}. The clumping of the stellar wind leads to significant variability in mass transfer and accretion rates in 4U 1700-377 \citep{Ducci2009}. The components of the system are also notable for their large masses, with the companion star having mass of $46 \pm 5$ \msol and the compact object having mass $1.96 \pm 0.19$ \msol \citep{Falanga2015}. This makes the companion to 4U 1700-377 one of the most massive known in a HMXB \citep{IslamPaul2016}. 

For these reasons 4U 1700-377 has the potential to provide a great deal of insight into unstable accretion processes seen in HMXBs, especially those which drive bright flares or arise due to mass exchange from an inhomogeneous stellar wind. Since the brightness of the source is highly variable due to frequent intense flares we benefit from being able to study a large body of observations across a great span of time. 

The presence of a CRSF in the spectrum of 4U 1700-377 has been an open question for more than two decades. Early observations suggested the presence of a cyclotron line at 37 keV \citep{Reynolds1999}, although more recent reports show that its appearance is model dependent \citep{Jiaswal2015a}, and confidence in this feature is fairly low \citep{Seifina2016}. Using a focused NuSTAR observation, \citet{Bala2020} was able to identify a wide, lower energy CRSF in the NuSTAR data. They report a confidence of $4-5 \sigma$ depending on the continuum model used. \citet{Xiao2024} has recently found the CRSF to be highly model dependent, a result which is supported by a more recent analysis of NuSTAR's focused observations by \citet{WO2026}. In addition to the possible CRSF, \citet{Bala2020} was able to identify a $7.5$ keV nickel emission line, which had not been previously reported. Leveraging 881 ks worth of serendipitous ``stray light" observations from NuSTAR detected over the course of seven years, we study the variable behavior of 4U 1700-377 in order to better understand the appearance of the newly reported high-energy emission line and CRSF. 

The Nuclear Spectroscopic Telescope Array (NuSTAR) was launched in 2013 as the first focusing X-ray observatory, designed to operate in the range of 3-79 keV \citep{Harrison2013}. Due to the telescope's long mast connecting the mirrors to the detectors, shrouding the optical path would be infeasible. Thus, NuSTAR is susceptible to ``stray light" intrusion onto the detectors when a bright source is $\sim1-4$ deg off axis from the primary focused target \citep{Madsen2017}. Observations with NuSTAR which have been ``contaminated" by stray light actually often present additional data for the contaminant source, which can be extracted into high-level science products \citep{Grefenstette2021}. For bright, hard X-ray sources near the galactic ridge (which are most likely to contaminate other focused observations), the \texttt{StrayCats} catalog can provide a significant amount of additional data, especially useful for tracking behavior across longer timescales. 4U 1700-377 is one such source which commonly appears as stray light on NuSTAR observations.

Confirming the detection of a CRSF would conclusively determine that the compact object in 4U 1700-377 is a neutron star, providing a key data point in the neutron star mass and equation of state problems, as well as provide another direct measurement of a neutron star magnetic field strength. Further measurements of fluorescence lines beyond Fe K$\alpha$ would also prove valuable for understanding the geometry of the stellar wind and other larger-scale features, such as an accretion wake \citep{Haberl1989, Goldstein2004}. In the following paper we present the analysis of 2 focused observations and 19 stray light observations of 4U 1700-377, discuss the consequences of our findings, and suggest next steps for making further insightful observations of this source. 

%%%%%%%%%%%%%%%%%%%%%%%%%%%%%%%%%%%%%%%%%%%%%%%%%%%%%%%%%%%%%%%%%%%%%%%%%%%
%%%%%%%%%%%%%%%%%%%%%%%%%%%%%%%%%%%%%%%%%%%%%%%%%%%%%%%%%%%%%%%%%%%%%%%%%%%
\section{Data Reduction} \label{sec:data-reduction}

\begin{figure}[t!]
    \centering
    \includegraphics[width=1\linewidth]{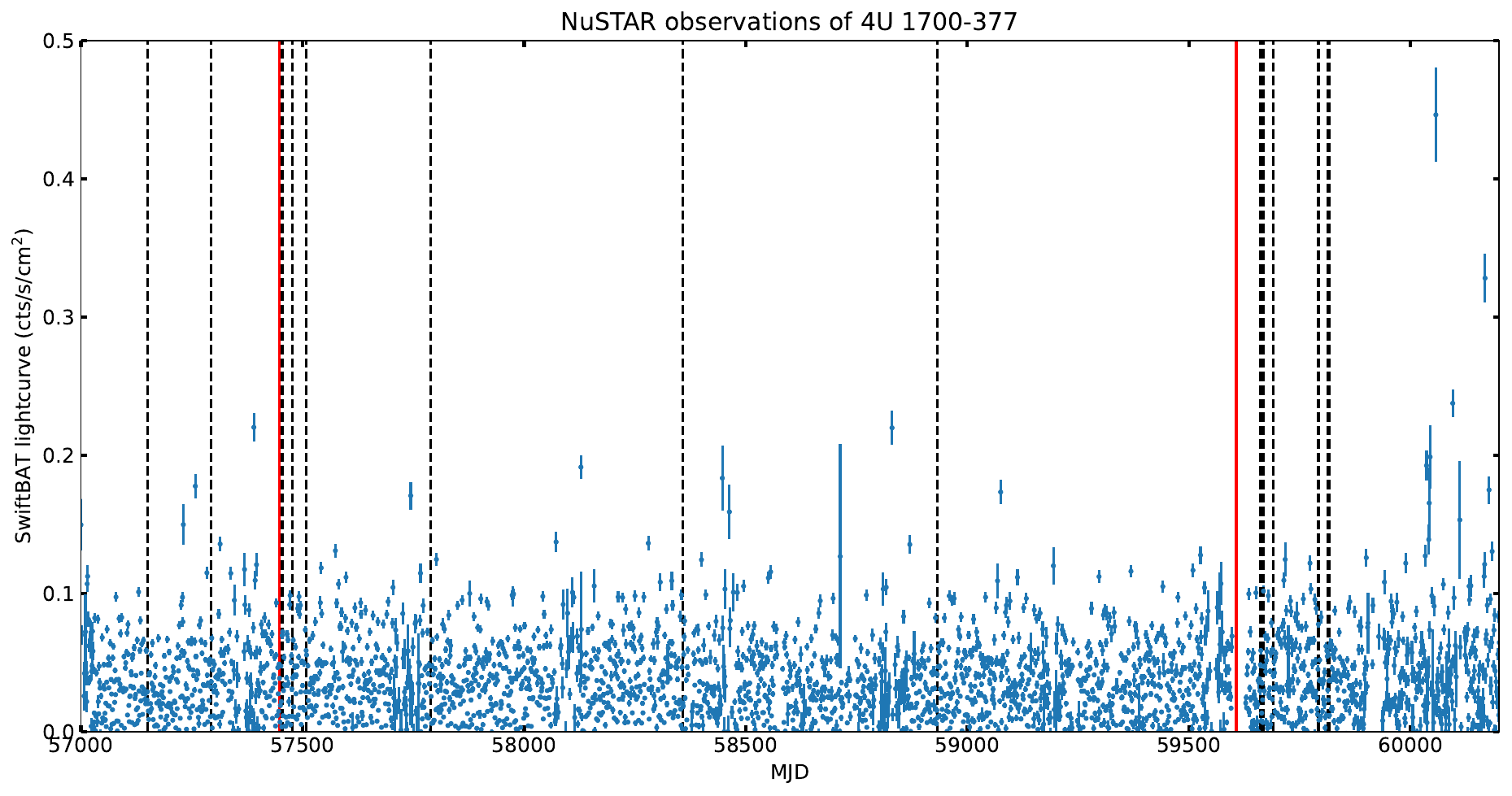}
    \caption{Timeline of all NuSTAR observations of 4U 1700-377. The two focused observations are indicated by solid vertical red lines, and the stray light detections are marked by dashed black lines.}
    \label{fig:SwiftBAT-lc-sl}
\end{figure}

In total (counting modules A and B separately) 4U 1700-377 appears 19 times in the \texttt{StrayCats} catalog, representing 881 ks of observation time over the course of seven years. Even with the significantly lower count rates that result from bypassing the mirrors, the amount of total stray light counts recorded for 4U 1700-377 is comparable to that of a full focused observation. The timeline of NuSTAR detections of 4U 1700-377 is shown in Figure \ref{fig:SwiftBAT-lc-sl}, along with the long term Swift/BAT light curve \citep{Barthelmy2005}. 

%% Observations timeline figure

The 19 stray light observations of 4U 1700-377 are reported in Table \ref{tab:obs}. As described in the table, some of these observations are deemed unsuitable for analysis, either due to excessively complicated contamination patterns, small stray light regions, or a focused target so bright that it illuminates the full detector. We also compare the results of the stray light data analysis with two focused observations described in the first two rows of Table \ref{tab:obs}. All data have been re-processed using the NuSTAR Data Analysis Software version 2.1.4 (\texttt{NuSTARDAS}) and \texttt{CalDB} version 20240229. 

For the two focused observations, a region of 50" centered on the target is selected as the source region, and another 50" region is selected off-target for the background. Spectra and response files are then extracted for both detectors and both observations using \texttt{nuproducts} version 0.3.3. 

Processing observation 30701023001 requires electing a modified ancillary response file (\texttt{nuA20100101v006.arf)} during the \texttt{nuproducts} call, in order to correctly account for a change in the effective area of FPMA below $\sim 7$ keV \citep{Madsen2020}. Focused observation 30101027002 is affected by the presence of a weak, soft ghost ray \citep{Madsen2017} directly on top of the source region on FPMA. When 4U 1700-377 is sufficiently bright, this ghost ray contributes relatively so few photons as to not affect the source spectrum (e.g., the FPMA/B normalization is well described by a constant factor of $\sim 0.97$). At lower flux states, however, the ghost ray adds a small amount of excess photons below 5 keV to the source spectrum. When time filtering for only these low flux states, we ignore data from 3-5 keV on FPMA as a means of filtering out the ghost ray. Above 5 keV, each detectors' spectra agree well, to within a constant factor of $\sim 0.93$.

Treatment of the stray light data is more complicated, as explained in \citet{Grefenstette2021}. Source regions for the stray light observations are based on those provided in the \texttt{StrayCats} catalog \citep{Ludlam2022}, and are shown for each observation in Appendix \ref{appendix:allprods}. Since X-rays beyond $\sim 30$ keV penetrate the aperture stops minimally attenuated \citep{Madsen2017} we cannot simply subtract a spectrum from a background region. Instead, a background is modeled based on known detector response behavior using \texttt{nuskybgd}\footnote{https://github.com/NuSTAR/nuskybgd-py} \citep{Wik2014}. 

Using the \texttt{nustar-gen-utils}\footnote{https://github.com/NuSTAR/nustar-gen-utils} Python wrappers created for NuSTAR's stray light, we extract source-region spectra for all of the stray light observations (see Fig.\ \ref{fig:all-spectra}).  For fitting the data, the focused and unfocused cosmic X-ray background components of \texttt{nuskybgd} are modeled based on the identified stray light region and left frozen. Spectral data from the stray light region at energies above 120 keV are used to constrain the instrument normalization component of the background, as described in \citet{Gulo2022}. Finally, although the instrumental lines component of the background generally dominates the spectrum at energies $\gtrsim 80$ keV \citep{Li2025}, 4U 1700-377's spectrum extends far enough into the hard X-rays that constraining this component by itself would not be reliable. Instead, the instrumental lines component of the background is left free and fit alongside the source spectrum. In all figures showing stray light spectra, the full source-region spectrum is plotted in the top panel in order to illustrate the fit of the background model, while residuals are only shown over energies where the source model fit was performed (between 3 keV and the energy at which the background begins to dominate, which for stray light will be impacted by factors such as the size of the illuminated region).

In all cases, spectra are binned using the binning scheme of \citet{KaastraBleeker2016}, with no minimum counts per bin requirement. For all observations and models we report unabsorbed flux in a range of both 3-12 keV and 3-50 keV. Fits and spectral analysis are performed using \texttt{Xspec} \citep{Arnaud1996} version 12.14.0. All errors reported in this paper are to 90\% confidence. In a few cases, a model can be fit with uncertainties on emission line centroid energy of less than 40 eV (one NuSTAR channel). For consistency, only statistical errors from the fit are reported, however narrow errors on line energies should be considered in the context of NuSTAR's instrumental resolution as well.

%% Stray light obs table
\begin{deluxetable*}{cclcccc}[ht]
% \tablewidth{0pt} %% Despite being reccomended in the AASTeX author guide, \tablewidth is actually known to not do anything for the deluxetable environment... :/
\tablecaption{NuSTAR observations of 4U 1700-377 \label{tab:obs}}
\tablehead{\colhead{ObsID} & \colhead{Module} & \colhead{Date} & \colhead{Start Time} & \colhead{Exposure} & \colhead{Area} & \colhead{Analyzed?} \\ 
\colhead{} & \colhead{} & \colhead{} & \colhead{(MJD)} & \colhead{(ks)} & \colhead{(cm2)} } 
\startdata
30101027002 & A,B & 2016 Mar 01 & 57448.45 & 38.02 & ... & Yes \\
30701023001 & A,B & 2022 Jan 27 & 59606.64 & 50.43 & ... & Yes \\ \hline
30001130002 & A & 2015 May 8 & 57150.22 & 81.27 & 2.7 & Yes \\
30001130002 & B & 2015 May 8 & 57150.22 & 80.92 & 1.6 & No\tablenotemark{a} \\
40111001002 & B & 2015 Sep 27 & 57292.83 & 49.48 & 4.3 & Yes \\
80001041002 & A & 2016 Mar 07 & 57454.08 & 43.29 & 3.2 & No\tablenotemark{b} \\
80001041002 & B & 2016 Mar 07 & 57454.08 & 43.21 & 3.2 & No\tablenotemark{b} \\
40111002002 & A & 2016 Mar 30 & 57477.38 & 57.35 & 4.4 & No\tablenotemark{c} \\
40111002002 & B & 2016 Mar 30 & 57477.38 & 57.14 & 4.3 & No\tablenotemark{c} \\
80202015002 & B & 2016 Apr 30 & 57508.48 & 38.96 & 4.1 & No\tablenotemark{b} \\
40201001002 & A & 2017 Feb 5 & 57789.38 & 62.78 & 2.9 & Yes \\
30401023002 & B & 2018 Aug 28 & 58358.64 & 93.07 & 3.2 & Yes \\
90601310002 & A & 2020 Mar 24 & 58932.21 & 41.33 & 2.5 & Yes \\
90601310002 & B & 2020 Mar 24 & 58932.21 & 41.03 & 0.8 & No\tablenotemark{a} \\
80702315002 & B & 2022 Mar 23 & 59661.16 & 14.98 & 2.1 & No\tablenotemark{b} \\
80702315004 & B & 2022 Mar 26  & 59664.58 & 18.17 & 1.4 & No\tablenotemark{b} \\
80702315006 & B & 2022 Mar 29 & 59668.00 & 15.58 & 1.7 & No\tablenotemark{b} \\
80802321002 & B & 2022 Apr 21 & 59690.20 & 17.75 & 2.1 & No\tablenotemark{b} \\
80802321005 & A & 2022 Jul 31 & 59791.85 & 15.89 & 5.8 & No\tablenotemark{b} \\
80801324002 & A & 2022 Aug 22 & 59813.04 & 27.63 & 6.1 & No\tablenotemark{b} \\
80801324004 & A & 2022 Aug 25 & 59816.66 & 81.85 & 5.7 & No\tablenotemark{b} \\ 
\enddata
\tablenotetext{a}{Area $< 2$ cm$^2$}
\tablenotetext{b}{Bright focused target}
\tablenotetext{c}{Extended focused target}
\end{deluxetable*}

\begin{figure}
    \centering
    \includegraphics[width=1\linewidth]{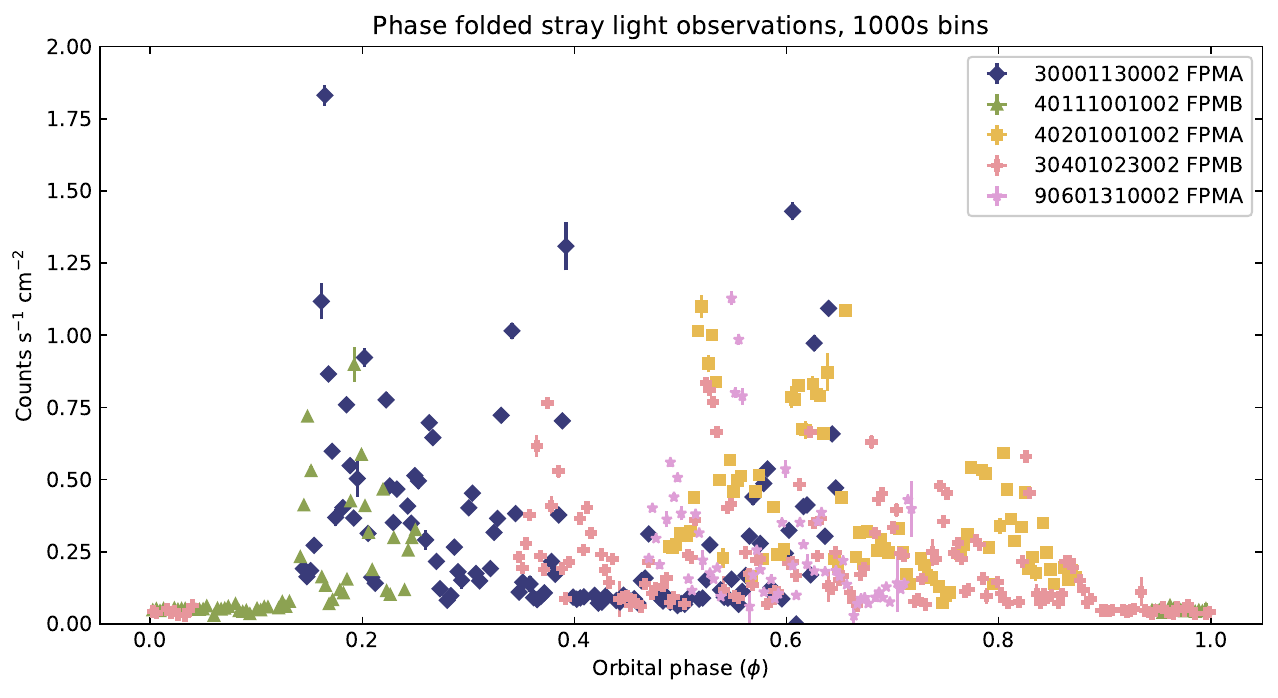}
    \caption{Light curves from every stray light observation of 4U 1700-377 plotted against phase of the binary orbit. In total, the stray light detections span the entire orbit, and demonstrate the high variability of the source brightness over the timescale of individual observations.}
    \label{fig:combo_lc_phi}
\end{figure}

For all observations, we extract light curves in the range of 3-20 keV. We also note that count rates for stray light observations appear lower (by a factor of 10-100) than a focused observation for the same flux of source photons, due to the significant decrease in effective-area when bypassing NuSTAR's mirrors. For this reason, stray light light curves are extracted in much longer, 1000 s bins (see Fig.\ \ref{fig:all-sl-lightcurves}). Using the orbital ephemeris from \citet{IslamPaul2016} we are able to determine orbital phase for each observation. A combined, phase-folded light curve is presented in Figure \ref{fig:combo_lc_phi} in order to demonstrate the extent to which the stray light observations span 4U 1700-377's binary orbit, as well as the source's extreme variability. 

%% Phase folded SL light curves

%%%%%%%%%%%%%%%%%%%%%%%%%%%%%%%%%%%%%%%%%%%%%%%%%%%%%%%%%%%%%%%%%%%%%%%%%%%
%%%%%%%%%%%%%%%%%%%%%%%%%%%%%%%%%%%%%%%%%%%%%%%%%%%%%%%%%%%%%%%%%%%%%%%%%%%
\section{Spectral Analysis} \label{sec:spectral-analysis}

While a variety of continuum models are commonly employed in modeling X-ray binaries like 4U 1700-377 \citep{Staubert2018}, it has been shown \citep{Bala2020, Xiao2024} that for this particular source, the appearance of certain spectral features (emission lines, CRSF) is often tied to the choice of continuum model. As such, as a first step in our analysis of the spectrum of 4U 1700-377, we determine a suitable continuum model which will ensure quality fits and consistency across observations. The specific choices of continuum model are described in the following subsection.

Regardless of the continuum model chosen, we include an absorption component describing the line-of-sight column density to the source. Each model also includes at least one Gaussian emission line component, to describe a prominent Fe K$\alpha$ line which is persistent in this source \citep[e.g.][]{GG2015}; in many cases, a second emission line (discussed in greater detail in Section \ref{subsec:emission-lines}) significantly improves the fit of the model. Additionally, a soft excess has previously been reported for 4U 1700-377 \citep{Boroson2003, Hickox2004, Rikame2024}, and while NuSTAR's 3 keV low energy limit is not ideal for constraining precise measurements of a soft excess, we do still find that a \texttt{bbody} component improves the model for all observations. This feature is detected at a confidence of $>3\sigma$ (as tested using Xspec's \texttt{simftest} routine) in both focused observations and most stray light observations, although the confidence is as low as 90\% for stray light spectra with more limited photon statistics (e.g.\ ObsID 90601310002). Finally, in order to co-model data from NuSTAR's two FPMs, a constant scaling factor is applied to account for normalization differences between the modules of a few percent.

\subsection{Selection of the Continuum Model} \label{subsec:continuum-models}

High-mass X-ray binary spectra are commonly modeled using either phenomenological, power law-based models or more physically-motivated analytical models \citep{Staubert2018}. Generally both of these types of models may be capable of producing good fits, although deriving physical insight requires care in how the model is applied: either by being aware of the compromises made in an empirical model, or by ensuring analytical models are representative of the source being observed. For modeling the focused observations of 4U 1700-377, we employ a choice of two empirical models and two analytical models, described in detail below.

\subsubsection{Empirical Models}

Power law-based models are commonly employed for modeling continua of HMXBs, based on their agreement with both the prediction of Comptonization as the primary contribution to the high energy continuum, and empirical agreement with observations \citep{Staubert2018}. Generally, these power laws are modified by a high energy exponential cutoff, equivalently parameterized through an electron temperature, $kT_e$, or a folding energy $E_{fold}$, depending on the model. This approach is analytically motivated \citep{Staubert2018} and provides a higher quality fit to most X-ray binary spectra. In order to account for the high-energy exponential cutoff seen in the spectrum of 4U 1700-377, we consider several modifications to a simple power law continuum: \texttt{highecut}, \texttt{cutoffpl}, and \texttt{FDcut} \citep{Tanaka1986}. All of these models adjust a power law continuum by a multiplicative factor proportional to $e^{-E}$. In the case of \texttt{highecut}, this modification is sharply introduced for energies above a certain $E_{cutoff}$; for \texttt{cutoffpl} and \texttt{FDcut}, this factor is applied across the entire energy range. Of these models, we find that only \texttt{highecut} is capable of producing high-quality fits to the spectra of 4U 1700-377. 

  It is important to consider the discontinuity in \texttt{highecut} which is introduced at $E_{cutoff}$ \citep{Staubert2018} which might mask excess emission. For the NuSTAR observations of 4U 1700-377 this cutoff generally occurs between 7 and 8 keV, coinciding with the emission features discussed below. This discontinuity can be smoothed, however, with the addition of a small Gaussian absorber \citep[e.g.][]{Coburn2002, Furst2013, Hemphill2019, Bala2020}. In our spectra, smoothing this discontinuity has no meaningful impact on the continuum model parameters or uncertainties, although it would allow a masked emission line (such as Ni K$\alpha$) to be added to the \texttt{highecut} models. Attempting to add back in this emission line does create degeneracies in the model which makes it difficult to constrain its energy \citep{Bala2020}, so for simplicity we fit \texttt{highecut} models with only one \texttt{gaus} profile, which accounts for Fe K$\alpha$.

In addition to the single power law models discussed above, we also consider the model \texttt{NPEX} \citep{Mihara1995}, which is the sum of a negative and positive index power law modified at all energies by an exponential cutoff. The inclusion of the positively-signed power law component in \texttt{NPEX} allows the model to better describe a Wien hump and thermal rollover expected for the spectrum of accreting pulsars \citep{Makishima1999}.

\subsubsection{Analytical Models}

Application of an analytical model to the spectrum of 4U 1700-377 is made complicated by the present uncertainty surrounding the nature of the compact object. While choosing analytical models necessarily requires assumptions about the physical system, we attempt to keep these assumptions consistent with existing literature, and caution that physical interpretations of the results of spectral fits with the analytical models are made in the context of uncertainty surrounding the true nature of 4U 1700-377's compact object. 

Presuming a blackbody source of seed photons, we apply the model \texttt{thcomp} \citep{Zdziarski2020}, which describes the Comptonization of these photons by thermal electrons. We find that Comptonization of a single, hot blackbody can reproduce the observed spectra of 4U 1700-377 quite well.

Additionally, while a lack of pulsations \citep{Dolan1980, Xiao2024} and uncertainty surrounding a CRSF \citep{Xiao2024, WO2026} leave no conclusive evidence for 4U 1700-377's compact object as a neutron star, other spectral indicators do still point towards a neutron star compact object \citep{Seifina2016, Chicharro2018}. As such, analytical models describing magnetized accreting pulsars have been applied to this source previously \citep{Bala2020}. We consider the model \texttt{COMPMAG} \citep{Farinelli2012}, which is based on numerical solutions to radiative transfer in a neutron star magnetosphere.

\subsubsection{Continuum Fits}

\begin{figure*}[h!]
    \centering
    \includegraphics[width=.44\linewidth]{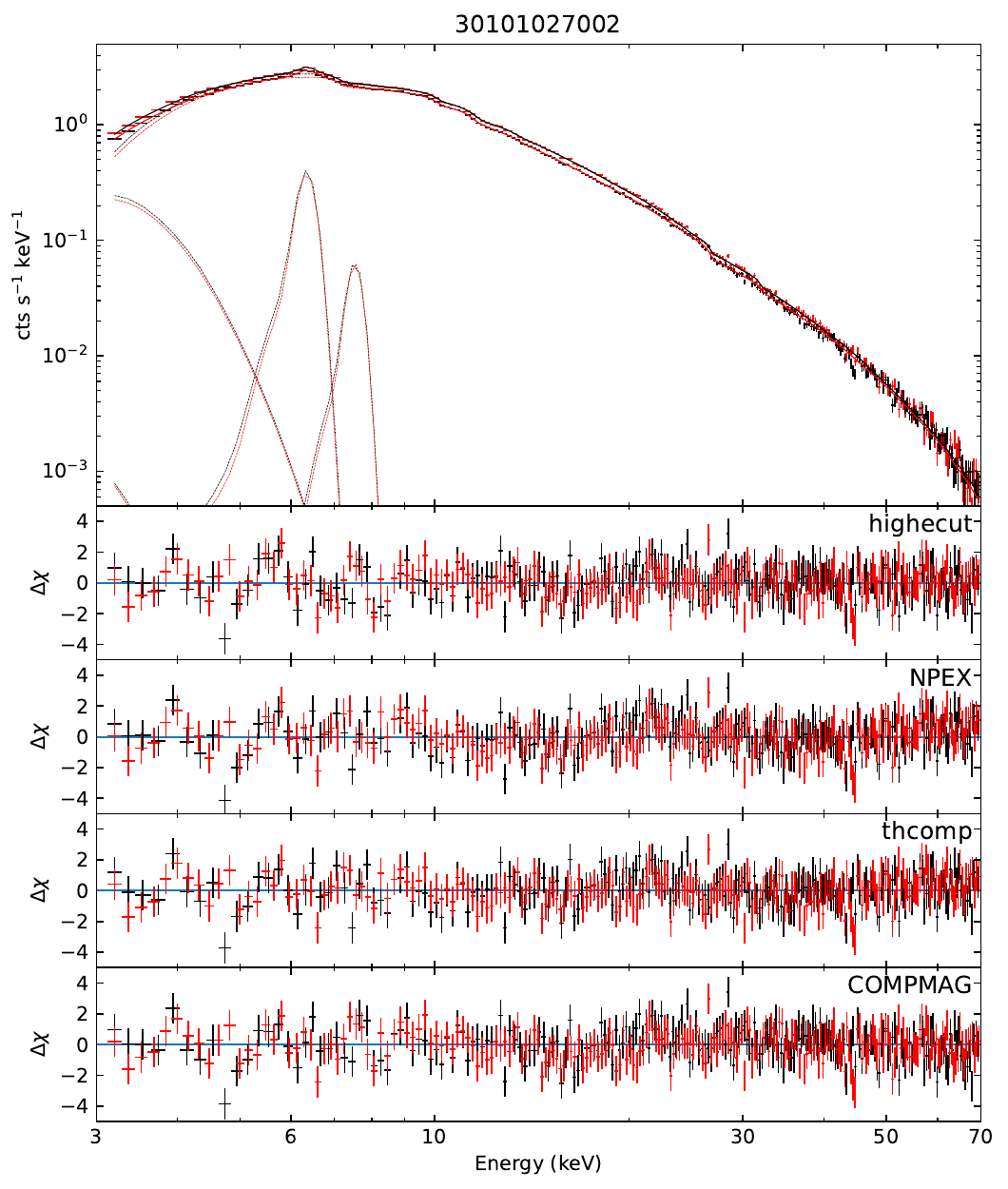}
    \includegraphics[width=.44\linewidth]{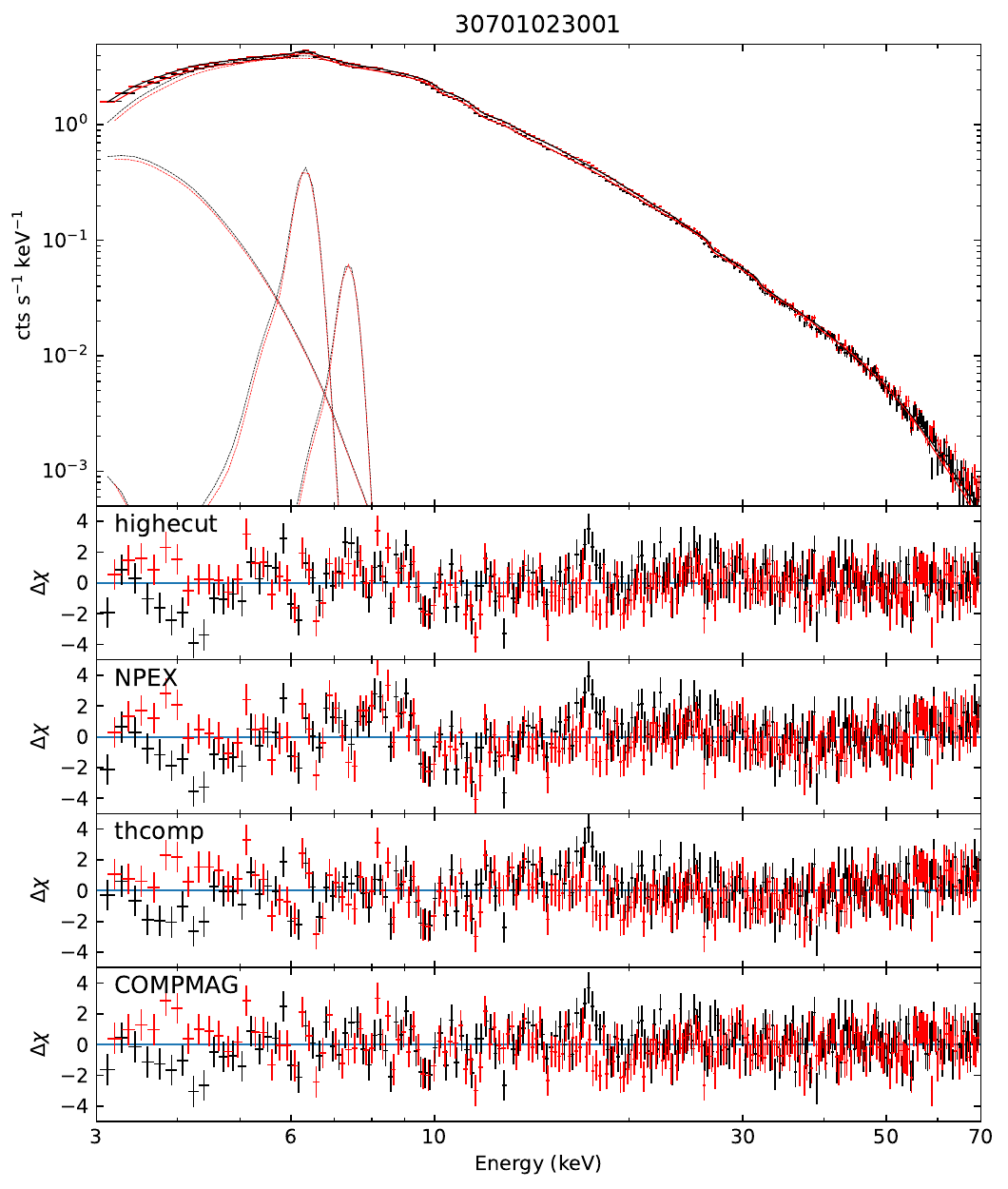}
    \caption{Selection of four continuum models which can be well-fit to the spectrum of 4U 1700-377. The top panel in each column shows the NuSTAR spectrum, model, and components; each panel below shows the $\Delta\chi$ residuals for \texttt{highecut}, \texttt{thcomp}, \texttt{NPEX}, and \texttt{COMPMAG} continuum models. A \texttt{highecut} continuum model yields the best fit to observation 30101027002, and the second best fit for 30701023001; the more complicated \texttt{COMPMAG} offers a slight improvement in reduced $\chi^2$ for the latter observation. All data and models are plotted such that FPMA appears in black, and FPMB in red. \label{fig:continuum-model-comparison}}
\end{figure*}

Parameters for the spectral fits with all four continuum models are reported in Table \ref{tab:focused-models-params}. In Figure \ref{fig:continuum-model-comparison} we show both focused spectra along with the residuals from the best fit of each of the above models. We find that \texttt{highecut}, while requiring the fewest free parameters, still provides the best fit for observation 30101027002 and the second-best fit for observation 30701023001 by $\chi^2/\textnormal{d.o.f}$. We note that this model is equivalent to the HCUT model employed in \citet{Bala2020}. This relative simplicity makes \texttt{highecut} a good choice as base continuum model for the stray light observations, which have much more limited signal-to-noise, and would risk being over-fit by a more elaborate continuum model. Overall, the fits reported for the focused observations generally agree with the models described in \citet[ObsID: 30101027002]{Bala2020} and \citet[ObsID: 30701023001]{WO2026}. 

For the two focused observations we find that absorption differs by around a factor of 3-4, with observation 30101027002 more highly obscured than 30701023001. Still, photon/spectral indices are generally consistent between the observations, around $\sim 1.28$ for \texttt{highecut}, $\sim 0.76$ for \texttt{NPEX}, and $\sim 1.84$ for \texttt{thcomp}. For the empirical models, folding energies of $\sim20$ keV (\texttt{highecut}) or $\sim10$ keV (\texttt{NPEX}) are consistent with the range generally seen in accreting pulsars \citep{White1983, Makishima1999, Filippova2005}.

Both focused observations show the presence of a ``soft excess" \citep{Hickox2004}, which we model as blackbody of temperature $\sim 0.3$ keV. Assuming a distance of $1.63 \pm 0.15$ kpc to 4U 1700-377 \citep{BJ2018, vdMeij2021} gives a total unabsorbed blackbody luminosity in the range of $\sim 3-300 \times 10^{36}$ ergs s$^{-1}$ for focused observation 30701023001 and between $ 5-40 \times 10^{36}$ ergs s$^{-1}$ for 30101027002, depending on the choice of continuum model. These luminosities indicate a total emitting surface area of a few hundreds of kilometers in radius; the empirical models find a range of radii between $100-1200$ km, while the analytical models favor a range of $20-170$ km. In either case, this suggests that the ``soft excess" observed in 4U 1700-377 is the result of the photoionized stellar wind in the vicinity of the compact object \citep{Mushtukov2022}.

The analytical models \texttt{COMPMAG} and \texttt{thcomp} both also describe a hotter blackbody of temperature $\sim 1.3$ keV, whose photons are Comptonized to form the high-energy continuum \citep{Farinelli2011, Zdziarski2020}. Both models indicate this hotter blackbody to have an extent of only a few (2-5) km in radius. Presuming a neutron star compact object, this hot blackbody would be best explained as the hot spot on the surface of the compact object, with the emission being up-scattered in the polar accretion column \citep{Mushtukov2022}.

%% Focused obs - model comparison spectra

%% Fit parameters for focused observations
\startlongtable
\begin{longrotatetable}
\begin{deluxetable}{lcccccccc} \label{tab:focused-models-params}
\tablecaption{Fit parameters for focused observations.}
\tablehead{\multicolumn{1}{c}{} & \multicolumn{2}{c}{highecut\tablenotemark{a}} & \multicolumn{2}{c}{NPEX\tablenotemark{b}} & \multicolumn{2}{c}{thcomp\tablenotemark{c}} & \multicolumn{2}{c}{COMPMAG\tablenotemark{d}}\\
\colhead{Param.} & \colhead{30101027002} & \colhead{30701023001} & \colhead{30101027002} & \colhead{30701023001} & \colhead{30101027002} & \colhead{30701023001} & \colhead{30101027002} & \colhead{30701023001} }
\startdata
$N_H\ (10^{22}$ cm$^{-2}$)    & 12.1$\pm 0.6$ & 6.6$\pm 0.3$ & 12.2$_{-0.5}^{+0.6}$ & $6.9_{-0.3}^{+0.4}$ & 13.3$_{-0.5}^{+1.1}$ & 6.6$_{-0.5}^{+0.4}$ & 10.2$_{-0.7}^{+0.6}$ & 5.1$_{-0.7}^{+0.6}$ \\
Photon Index                & 1.25$\pm 0.03$ & 1.29$\pm 0.01$ & 0.74$\pm 0.02$ & 0.77$\pm 0.01$ & 1.84$\pm 0.01$ & $1.838_{-0.004}^{+0.007}$  & ... & ... \\
$E_{cutoff}$ (keV)          & 7.4$\pm 0.2$ & 7.3$\pm 0.1$ & ... & ... & ... & ... & ... & ... \\
$E_{fold}$ (keV)            & 23.2$\pm 0.7$ & 24.1$\pm 0.5$ & 9.8$_{-0.2}^{+0.3}$ & $9.5\pm0.2$ & ... & ...  & ... & ... \\ %% efold now including npex
$kT_e$ (keV)                & ... & ... & ... & ... & $10.9\pm0.1$ & $10.9\pm0.1$  & 0.871$_{-0.092}^{+0.004}$ & $4.3_{-0.5}^{+0.9}$ \\ % this row combines compmag and thcomp electron temps
Cov.\ Frac.\                & ... & ... & ... & ... & $> 0.99$ & $>0.99$  & ... & ... \\
kT$_{bb}$ (keV)             & ... & ... & ... & ... & 1.33$_{-0.02}^{+0.03}$ & 1.31$_{-0.02}^{+0.01}$  & 1.26$\pm 0.03$ & 1.33$_{-0.03}^{+0.02}$ \\ % this row combines compmag and thcomp seed bbody temps
$\tau$                      & ... & ... & ... & ... & ... & ... & 0.46$_{-0.02}^{+0.03}$ & 0.39$_{-0.03}^{+0.01}$ \\
$r_0$                       & ... & ... & ... & ... & ... & ... & 0.171$_{-0.003}^{+0.001}$ & $0.25_{-0.03}^{+0.04}$ \\
A                           & ... & ... & ... & ... & ... & ... & 0.44$\pm 0.1$ & $0.5860_{-0.0003}^{+0.0001}$ \\
kT (bbody, keV)             & 0.26$\pm 0.3$ & $0.23\pm0.03$ & 0.27$_{-0.03}^{+0.02}$ & $0.28\pm0.02$ & 0.43$_{-0.02}^{+0.01}$ & $0.43\pm0.02$ & 0.29$_{-0.02}^{+0.01}$ & $0.35_{-0.04}^{+0.03}$ \\
$E_{Fe K\alpha}$ (keV)      & 6.34$\pm 0.01$ & 6.32$\pm 0.01$ & 6.35$\pm 0.01$ & $6.33\pm 0.01$ & 6.34$\pm 0.01$ & 6.32$\pm 0.01$ & 6.35$\pm 0.01$ & $6.33\pm0.01$ \\ 
$\sigma_{Fe K\alpha}$ (eV)  & 100\tablenotemark{f} & 100\tablenotemark{f} & 100\tablenotemark{f} & 100\tablenotemark{f} & 100\tablenotemark{f} & 100\tablenotemark{f} & 100\tablenotemark{f} & 100\tablenotemark{f} \\
$E_{Line2}$ (keV)           & ... & ... & 7.59$_{-0.08}^{+0.07}$ & $7.42_{-0.06}^{+0.05}$ & $7.59\pm0.08$ & 7.37$\pm 0.08$ & 6.37$_{-0.97}^{+0.07}$ & $7.2_{-0.2}^{+0.2}$ \\
$\sigma_{Line2}$ (eV)       & ... & ... & 100\tablenotemark{f} & 100\tablenotemark{f} & 100\tablenotemark{f} & 100\tablenotemark{f} & 800$_{-200}^{+300}$ & 100\tablenotemark{f} \\
Const. (FPMB)               & 0.972$\pm 0.002$ & 1.048$\pm 0.002$ & 0.972$\pm 0.002$ & $1.048 \pm 0.02$ & 0.972$\pm 0.002$ & 1.048$\pm 0.002$ & 0.972$_{-0.008}^{+0.006}$ & $1.048\pm0.002$ \\ \hline
$F_{12}$\tablenotemark{g} & $1.18 \pm 0.03$ & $2.49 \pm 0.03$       & $1.19^{+0.03}_{-0.02}$ & $2.52 \pm 0.03$       & $1.26^{+0.03}_{-0.04}$ & $2.50 \pm 0.05$             & $1.03 \pm 0.03$       & $2.38 \pm 0.02$ \\
$F_{50}$\tablenotemark{h} & $2.67 \pm 0.03$ & $5.59 \pm 0.03$ & $2.68 \pm 0.03$       & $5.64^{+0.04}_{-0.03}$ & $2.74^{+0.04}_{-0.03}$ & $5.61 \pm 0.06$ & $2.52^{+0.03}_{-0.04}$ & $5.46^{+0.06}_{-0.07}$ \\
\hline 
$\chi^2$ / d.o.f.           & 447.38 / 443 & 658.01 / 460 & 479.09 / 441 & 780.29 / 458 & 503.27 / 440 & 680.32 / 457 & 451.78 / 438 & 592.28 / 456 \\
\enddata
\tablenotetext{a}{In Xspec notation: \texttt{const*phabs(highecut*pow+bbody+gaus)}. }
\tablenotetext{b}{\texttt{const*phabs(NPEX+bbody+gaus+gaus)}. }
\tablenotetext{c}{\texttt{const*phabs(thcomp*bbody+bbody+gaus+gaus)}. }
\tablenotetext{d}{\texttt{const*phabs(COMPMAG+bbody+gaus+gaus)}. }
\tablenotetext{f}{Parameter has been frozen.}
\tablenotetext{g}{3-12 keV unabsorbed flux, $\times 10^{-9}$ ergs cm$^{-2}$ s$^{-1}$}
\tablenotetext{h}{3-50 keV unabsorbed flux, $\times 10^{-9}$ ergs cm$^{-2}$ s$^{-1}$}
\end{deluxetable}
\end{longrotatetable}

The results of the fits to the stray light data are reported in Table \ref{tab:all-sl-params}, and are generally consistent with the parameters found for the focused observations. For the stray light observations, absorption also varies by up to a factor of four. Photon indices are generally consistent around $\sim 1.2-1.3$. For the high energy cutoff, $E_{cutoff}$ ranges from 7-9 keV, with a folding energy of 20-30 keV. A similar blackbody of temperature $\sim 0.3$ keV is also found in all of the stray light observations, and would again correspond to photoionization of the stellar wind. Spectra with these fits are reported in Appendix \ref{appendix:allprods} along with fit parameters, light curves, detector images, and extraction regions. The remainder of the spectral analysis presented in this paper discusses the variability of 4U 1700-377, as seen across our large baseline of observations. 

\subsection{Emission Line Structures} \label{subsec:emission-lines}

A prominent Fe K$\alpha$ line has been commonly reported for this source, and indeed we find this emission line appears strongly in every NuSTAR observation of 4U 1700-377. Some previous works have also identified emission features above 6.4 keV in the source spectrum. In particular, using data from XMM Newton showing eclipse of the compact object, \citet{vanderMeer2005} found evidence for higher ionization states of iron, in the form of a second Fe K$\alpha$ line around 6.7 keV, as well as an Fe K$\beta$ line around 7.1 keV. More recently, \citet{Bala2020} used data from NuSTAR (ObsID: 30101027002) to identify a 7.5 keV emission line, which they attribute to Ni K$\alpha$. 

To study the emission line structure in the spectra of 4U 1700-377, we employ two approaches for avoiding interference from the continuum model. The first approach is to consider a limited data range (5-10 keV) such that a high-energy exponential cutoff is not necessary to fit the continuum, as was done in \citet{Bala2020}. For our second approach, we consider continuum model \texttt{thcomp}, which allows for high quality fits to the full continuum, without introducing a discontinuity in the model. 

\begin{figure*}
    \centering
    \fig{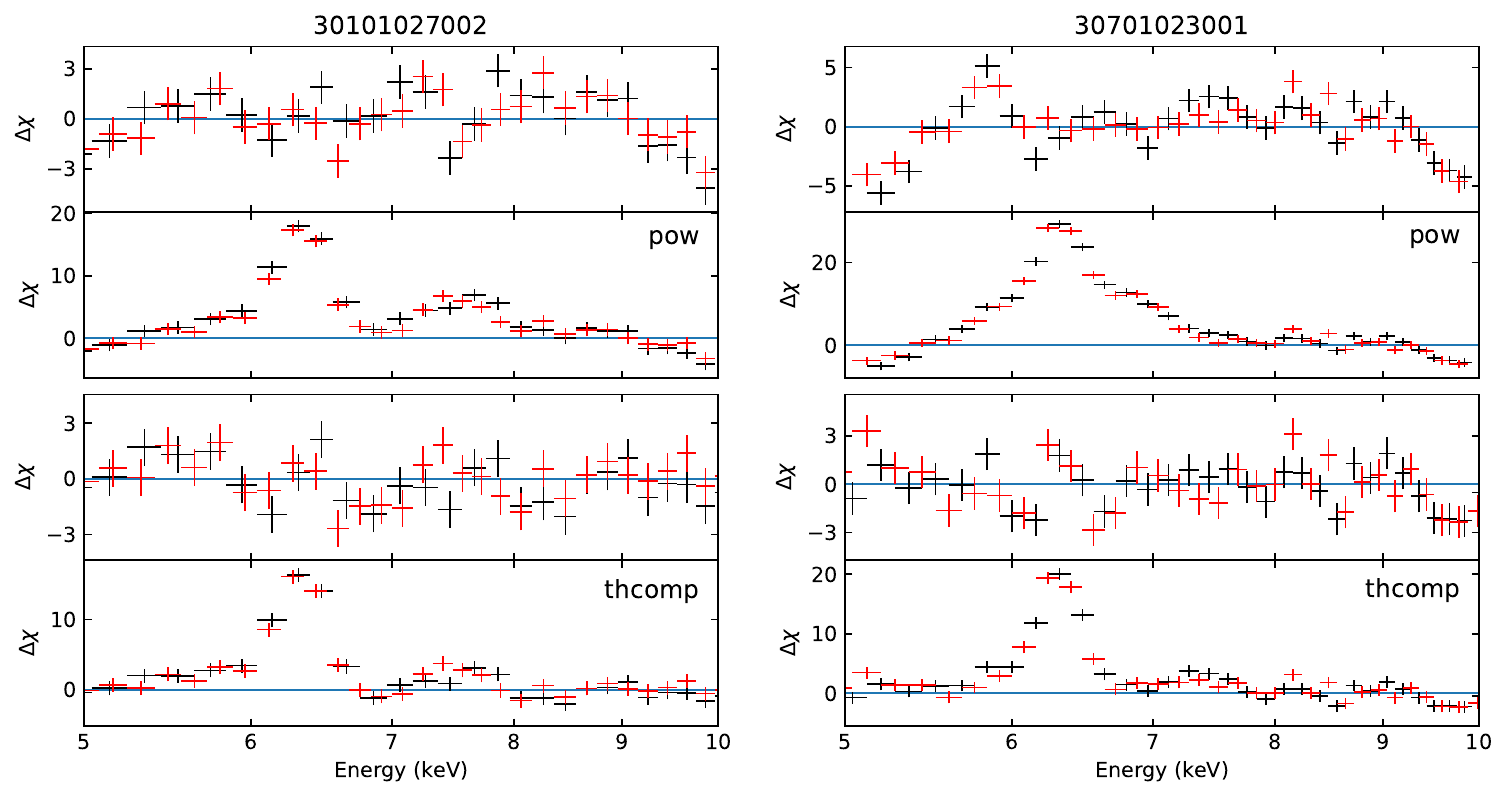}{0.8\textwidth}{}
    \caption{Best fit residuals and iron line profiles for focused NuSTAR observations 301010127002 (left) and 30701023001 (right). Top plots use \texttt{pow} as the continuum model; the bottom plots use \texttt{thcomp}. In all plots the top panels subplot shows $\Delta \chi$ residuals of the best-fit model, and the panels below show the $\Delta \chi$ when the emission line normalizations are set to 0.}
    \label{fig:focused-iron}
\end{figure*}

Applying these methods to the focused data shows some evidence for different states of high energy emission between 6.4 and 8 keV (Figure \ref{fig:focused-iron}). For observation 30101027002 we confirm the detection from \citet{Bala2020} of a line around $\sim 7.5$ keV. When the continuum is modeled with a simple power law, this line appears at $7.58 \pm 0.05$ keV, and for the \texttt{thcomp} continuum we measure the line at $7.59 \pm 0.08$ keV. The fits to focused observation 30701023001 are less clear, however. The simple power law continuum show no sign of Ni fluorescence, and instead favors a second emission line at $6.90 \pm 0.03$ keV. For the \texttt{thcomp} continuum, however, it is possible to fit an emission feature similar to the Ni K$\alpha$ line from the other observation, at $7.37\pm0.08$ keV. Even a third possibility, a two-component model of Fe K$\alpha$ consisting of a wide ($\sigma \sim 800$ eV) and narrow (unresolved) Gaussian \citep[more similar to][]{GG2015} ends up as the best description of the \texttt{COMPMAG} spectrum for observation 30101027002.

The NuSTAR data are generally confident in the detection of \textit{whichever} line is best fit to on given model and spectrum (\texttt{simftest} indicates $> 3\sigma$ confidence in any given line), suggesting that some, likely variable, excess emission in the region of Fe K$\alpha$ is present in the spectra of 4U 1700-377. The inability to constrain a cohesive understanding of these lines for a given observation, however, is a result of NuSTAR lacking sufficient spectral resolution to study these lines in detail. 

For the stray light observations, we repeat these approaches to test for high energy emission features. All of the observations indicate a slight improvement in the fit when another emission line is placed between 6.4-8 keV, however limited signal in the data means the second line is too weak and poorly constrained for analysis. 

%% Focused observations iron line residuals

\subsection{Investigation of an Absorption Feature} \label{subsec:absorption-feature}

The presence of a CRSF at either $\sim 16,\ 30,\ \textnormal{or}\ 50$ keV, as has been variously reported in the past \citep{Reynolds1999, Jiaswal2015a, Seifina2016, Bala2020}, has been rigorously investigated most recently by applying a variety of continuum models to Insight-HXMT data \citep{Xiao2024} and the focused NuSTAR data \citep{WO2026}. These studies have found that indications of a CRSF in the spectrum of 4U 1700-377 are model dependent and often low confidence. To compliment the analysis of the focused NuSTAR data, we also provide an assessment of possible absorption features seen in the stray light data.

\begin{figure}[h!]
    \centering
    \gridline{
    \fig{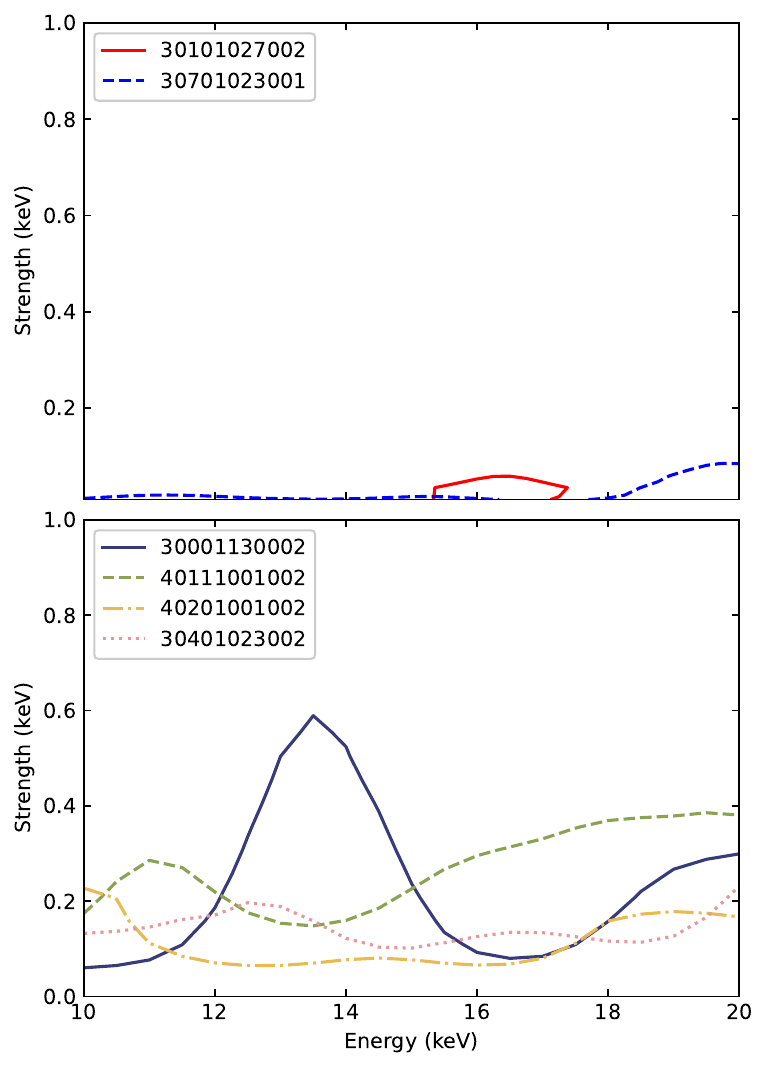}{0.4\textwidth}{}
    }
    \caption{90\% confidence intervals for a test absorption feature in the \texttt{highecut} continuum model, for both the focused (top) and stray light (bottom) observations. Open contours mean the \texttt{gabs} component is consistent with a strength of 0 keV. The contours shown here are generated for a width of 1 keV to illustrate a feature in observation 30001130002 (solid blue line) which is capable of being fit with a relatively stronger absorption feature. Still, this feature is consistent with zero-strength even at relatively low confidence, meaning its significance as part of the model is negligible.}
    \label{fig:highecut-contours}
\end{figure}

A blind search for absorption features above 10 keV does not strongly indicate any need to include a CRSF in our continuum models. Still, we test a few specific configurations of an absorption component (modeled by \texttt{gabs}) in the range of 10-20 keV, to test what strengths of absorption are compatible with the observed spectra. To do this, we check the fit statistic for each model when an absorption feature is set to a particular energy between 10-20 keV, and a particular strength between 0 and 1 keV. For a given run of this test, the width of the feature is frozen to a reasonable value, between 1-4 keV (to avoid a degeneracy between \texttt{gabs}' width and strength). Using the \texttt{highecut} continuum model, we generate sets of 90\% confidence intervals, shown in Figure \ref{fig:highecut-contours}.

For the stray light observations, lower signal-to-noise in the spectra leads to larger confidence intervals when testing for an absorption feature, compared to the focused observations. In agreement with the focused data, none of the stray light observations can constrain a non-zero absorption feature in the range of 10-20 keV. One observation (30001130002, the solid blue line in Fig.\ \ref{fig:highecut-contours}) can be fit with an absorption feature of relatively greater strength than the other observations, but the extent to which this improves the fit is not sufficient to claim this as a detection.

%% highecut contours

\subsection{Non-Flaring States} \label{subsec:time-filtered-low}

4U 1700-377's light curve shows significant variation in the form of frequent, intense, short-term flares on timescales of just 100s of seconds, as seen in Figure \ref{fig:combo_lc_phi}, and in Figure \ref{fig:all-sl-lightcurves}. Ignoring eclipse of the compact object, the source's count rate measured by NuSTAR still reliably varies by a factor of four or more, often multiple times in a single observation. In general this variability is well explained by simple Bondi-Hoyle accretion of a clumped stellar wind, as shown by \citet{Ducci2009}, who were able to reproduce observed flare luminosities, durations, and quantity. On several occasions, however, these flares can be seen to stop for an extended period of time, on the order of 10s of ks. In particular, we observe this extended pause in 4U 1700-377's flares in focused observation 30101027002 for $\sim 40$ks and in stray light observation 30001130002 for $\sim 60$ks. Studying these low states provides important insight into the accretion processes in 4U 1700-377, especially when in comparison to the source's flares. 

\begin{figure}
    \centering
    \gridline{
    \fig{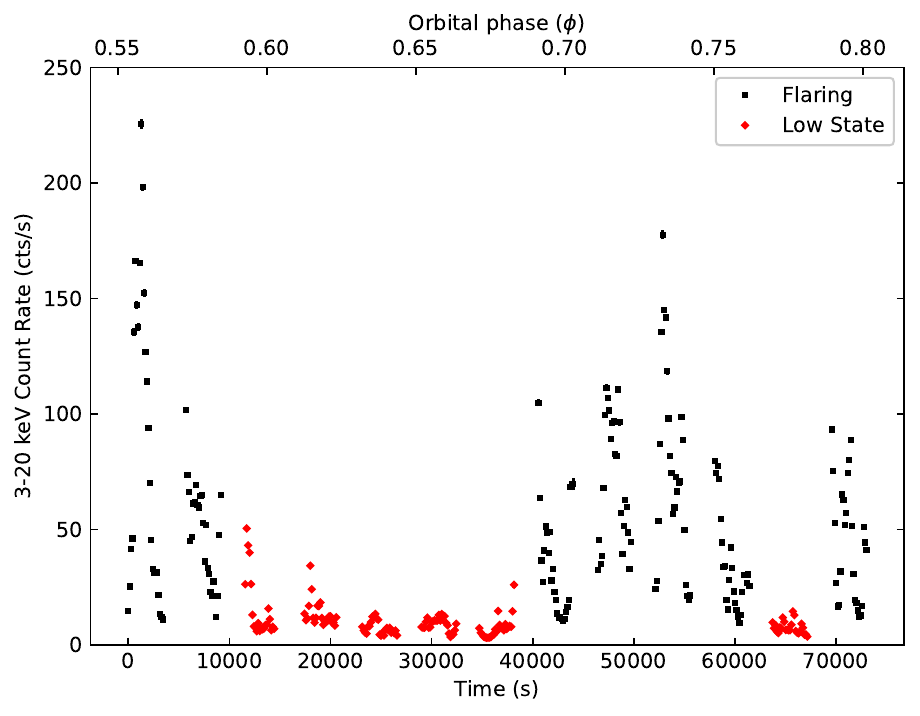}{0.4\textwidth}{}
    }
    \gridline{
    \fig{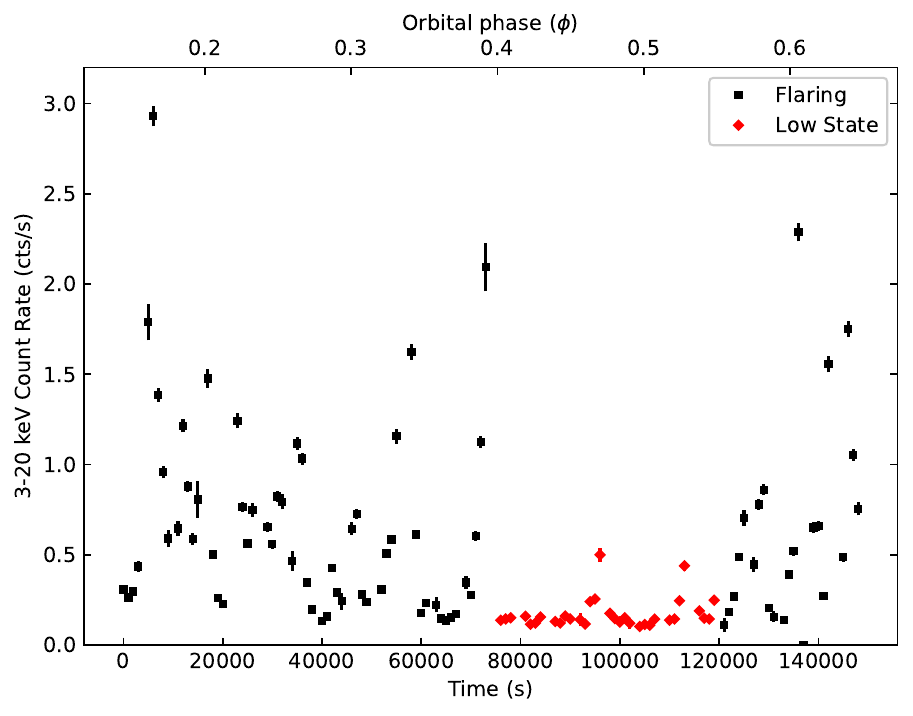}{0.4\textwidth}{}
    }
    \caption{Light curves for focused observation 30101027002 and stray light observation 30001130002, which both show extended low flux states. Times filtered for separate analysis are denoted black, for the flaring state and red for the low state.}
    \label{fig:low-lightcurves}
\end{figure}

In order to further study the behavior of the source during these low phases, we filter our spectra to select each observation's extended low count state. Time selections are made visually, using NuSTAR's Earth occultations as natural delimiters, as shown in Figure \ref{fig:low-lightcurves}. The identified bins represent intervals of one NuSTAR orbit whose averaged count rate $\pm$ one standard deviation is below that of the averaged light curve. We find that for the stray light observation the luminosity drops from $\sim 9 \times 10^{35}$ ergs s$^{-1}$ during the flaring state to $\sim 1.5 \times 10^{35}$ ergs s$^{-1}$ for the persistent emission. In the focused observation, source luminosity drops from $\sim 6.4 \times 10^{35}$ ergs s$^{-1}$ in the flaring state to $\sim 1.1 \times 10^{35}$ ergs s$^{-1}$ during the low state. 

%% Flare/low lightcurves for both obs

\begin{figure*}[h!]
    \centering
    \includegraphics[width=0.44\linewidth]{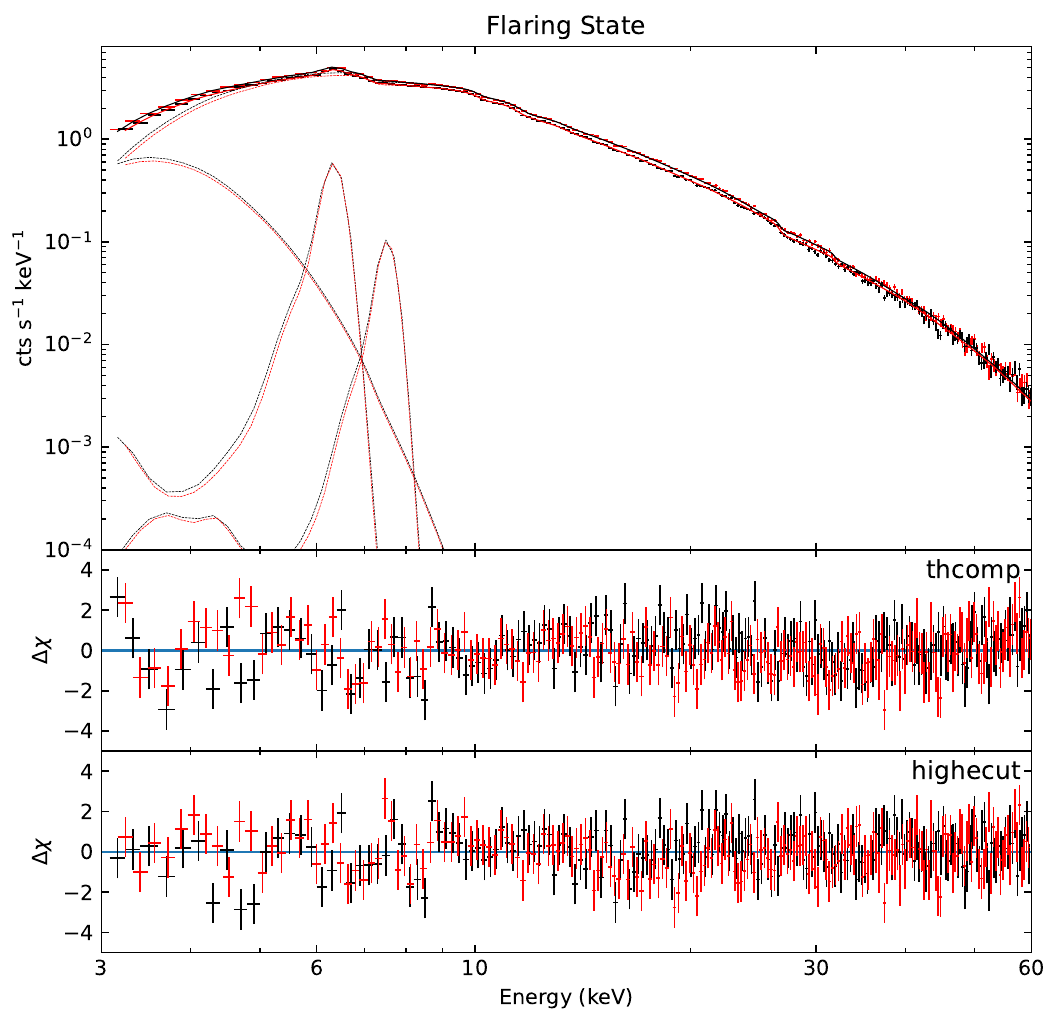}
    \includegraphics[width=0.44\linewidth]{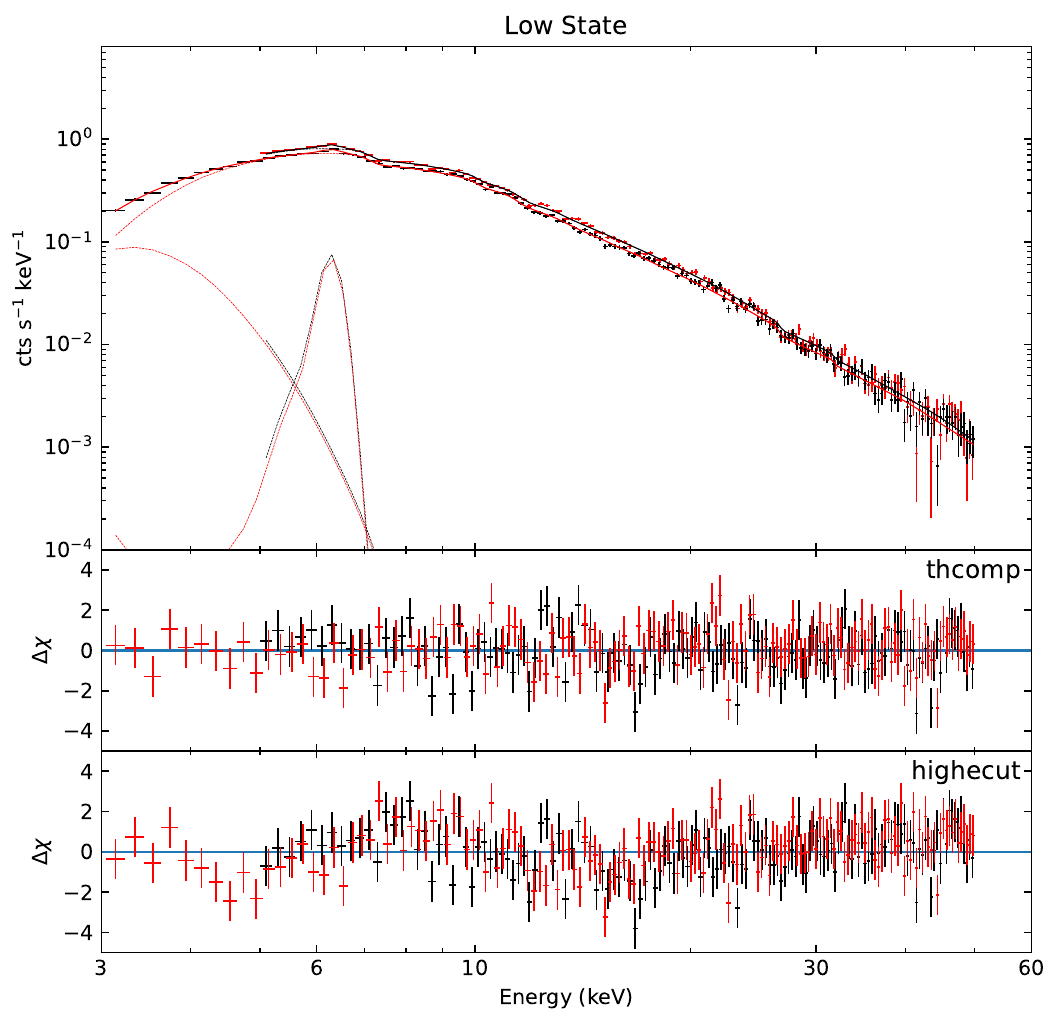}
    \caption{Flaring and persistent spectra for focused observation 30101027002. Panels (a.i) and (b.i) show $\chi^2$ residuals for the best fit \texttt{thcomp} model. Panels (a.ii) and (b.ii) are for a \texttt{highecut} model. Data points in black are from FPMA, red are from FPMB. In the low state, data from FPMA is ignored below 5 keV due to the presence of a weak ghost ray overlapping with the source region.}
    \label{fig:low-focused-spectra}
\end{figure*}

For the focused observation we observe changes in the source continuum between the low and flaring states, shown in Figure \ref{fig:low-focused-spectra}. Using the \texttt{highecut} continuum model, we find the photon index is significantly softer during the low state ($\sim 1.07$ while flaring to $\sim 1.79$ persistent). We also notice a slight increase in the absorbing column density during the low state, from $\sim 10^{23}$ cm$^{-2}$ to $\sim 1.5 \times 10^{23}$ cm$^{-2}$. The low state spectra also fit with higher cutoff and folding energies, compared to the flaring spectra. Blackbody temperatures and iron K$\alpha$ emission line parameters remain consistent between the two states. 

In the case of the focused low-state spectrum the fit with just a \texttt{highecut} and \texttt{bbody} model is slightly poor ($\chi^2 = 388.13 / 292$ d.o.f.), with noticeable negative residuals between 10-20 keV. While adding an absorption feature around 15 keV does tend to improve the fit statistic, the width of this feature is completely unconstrained. It is more likely that this negative residual is a byproduct of an incomplete description of the continuum. As such, we also consider a \texttt{thcomp} model for both states. For the low state, this model does allow for an improved fit statistic of $\chi^2/$d.o.f.\ $= 303.96/291$. The \texttt{thcomp} model also provides a good fit to the flaring spectrum when an additional emission line is modeled at 7.5 keV. This line is detected at a confidence of $\sim 3.3 \sigma$ (999/1000 simulated comparisons). We are not, however, able to confidently detect this line in the lower flux state, most likely a result of the more limited photon statistics at these low count rates.

Comparing \texttt{thcomp} models, we find the absorbing column density is consistent around $\sim 13 \times 10^{22}$. The low energy power law index is significantly harder during the flaring state, at $\Gamma_\tau \sim 1.8$, compared to a persistent photon index of $\sim 2.2$. The electron temperature of the Comptonizing material is lower during the flaring state, at around 10 keV, compared to the low-state's 20 keV. In both states, the fraction of seed photons upscattered is consistent with 1. Parameters for both states and both continuum models are reported in Table \ref{tab:low-focused-models}. For both the flaring and low states of the focused observation, continuum model \texttt{thcomp} does not indicate the presence of any absorption feature.

%% 30101027002 flare/low spectra

\begin{deluxetable*}{lcccc}[h!] \label{tab:low-focused-models}
\tablecaption{Flaring and persistent states of the focused observation 30101027002.}
\tablehead{
\multicolumn{1}{c}{} & \multicolumn{2}{c}{highecut\tablenotemark{a}} & \multicolumn{2}{c}{thcomp\tablenotemark{b}} \\
\colhead{Param.} & \colhead{Flaring} & \colhead{Persistent} & \colhead{Flaring} & \colhead{Persistent} 
} 
\startdata
$N_H\ (10^{22}$ cm$^{-2}$) & 10.5$\pm0.5$ & 15.3$_{-1.1}^{+1.2}$ & 13.8$_{-0.9}^{+0.7}$ & 13$_{-1}^{+2}$ \\
Photon Index & 1.07$\pm0.03$ & 1.79$_{-0.06}^{+0.05}$ & $1.82\pm0.01$ & $2.15_{-0.08}^{+0.02}$ \\
$E_{cutoff}$ (keV) & 6.6$\pm0.2$ & 7.9$\pm0.3$ & ... & ... \\
$E_{fold}$ (keV) & 20.0$_{-0.7}^{+0.5}$ & 31$\pm3$ & ... & ... \\
$kT_e$ (keV) & ... & ... & $10.5\pm0.2$ & 20$_{-5}^{+4}$ \\
Cov.\ Frac.\ & ... & ... & $> 0.99$ & $>0.85$ \\
kT$_{bb}$ (keV) & ... & ... & 1.38$\pm0.03$ & $1.3\pm0.1$ \\
kT (bbody, keV) & 0.24$\pm0.04$ & 0.19$_{-0.05}^{+0.04}$ & 0.43$\pm0.01$ & $0.36\pm0.06$ \\
$E_{Fe K\alpha}$ (keV) & 6.35$_{-0.01}^{+0.02}$ & 6.31$\pm0.05$ & 6.34$\pm0.01$ & 6.30$\pm0.05$ \\
$\sigma_{Fe K\alpha}$ (eV) & 100 \tablenotemark{f} & 100 \tablenotemark{f} & 100 \tablenotemark{f} & 100 \tablenotemark{f} \\
$E_{Ni K\alpha}$ (keV) & ... & ... & 7.5386$_{-0.0002}^{+0.0004}$ & ... \\
$\sigma_{Ni K\alpha}$ (eV) & ... & ... & 100 \tablenotemark{f} & ... \\
Const.  & 0.974$\pm0.003$ & 0.930$\pm0.007$ & 0.974$\pm0.003$ & 0.933$_{-0.007}^{+0.008}$ \\ \hline
$F_{12}$\tablenotemark{g} & $1.79 \pm 0.03$       & $0.37 \pm 0.02$ & $2.00^{+0.06}_{-0.07}$ & $0.34^{+0.03}_{-0.02}$ \\
$F_{50}$\tablenotemark{h} & $4.26^{+0.04}_{-0.03}$ & $0.65 \pm 0.02$ & $4.50^{+0.07}_{-0.03}$ & $0.62^{+0.03}_{-0.02}$ \\
\hline  
$\chi^2$ / d.o.f. & 446.50 / 441 & 388.13 / 292 & 494.34 / 438 & 303.96 / 291 \\
\enddata
\tablenotetext{a}{In Xspec notation: \texttt{const*phabs(highecut*pow+bbody+gaus)}. }
\tablenotetext{b}{\texttt{const*phabs(thcomp*bbody+bbody+gaus+gaus)}. }
\tablenotetext{f}{Parameter has been frozen.}
\tablenotetext{g}{3-12 keV unabsorbed flux, $\times 10^{-9}$ ergs cm$^{-2}$ s$^{-1}$}
\tablenotetext{h}{3-50 keV unabsorbed flux, $\times 10^{-9}$ ergs cm$^{-2}$ s$^{-1}$}
\end{deluxetable*}

\begin{figure*}
    \centering
    \gridline{
    \fig{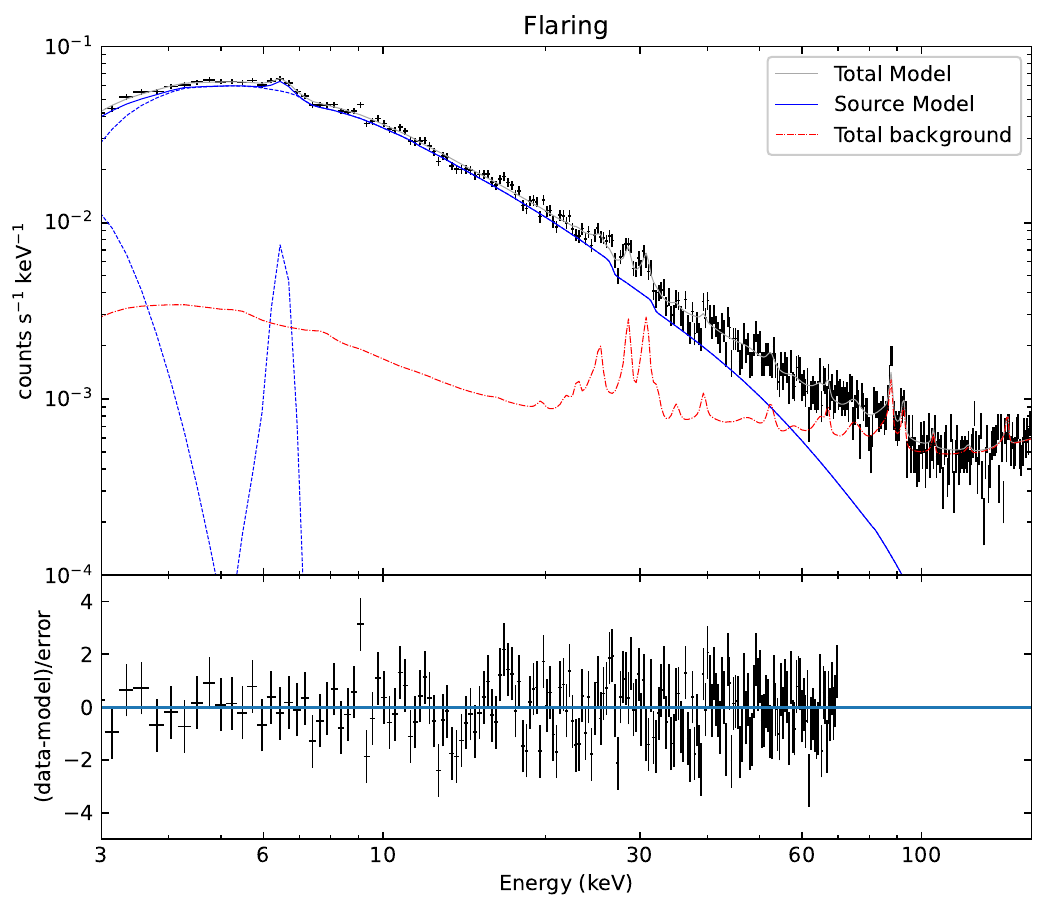}{0.46\textwidth}{}
    \fig{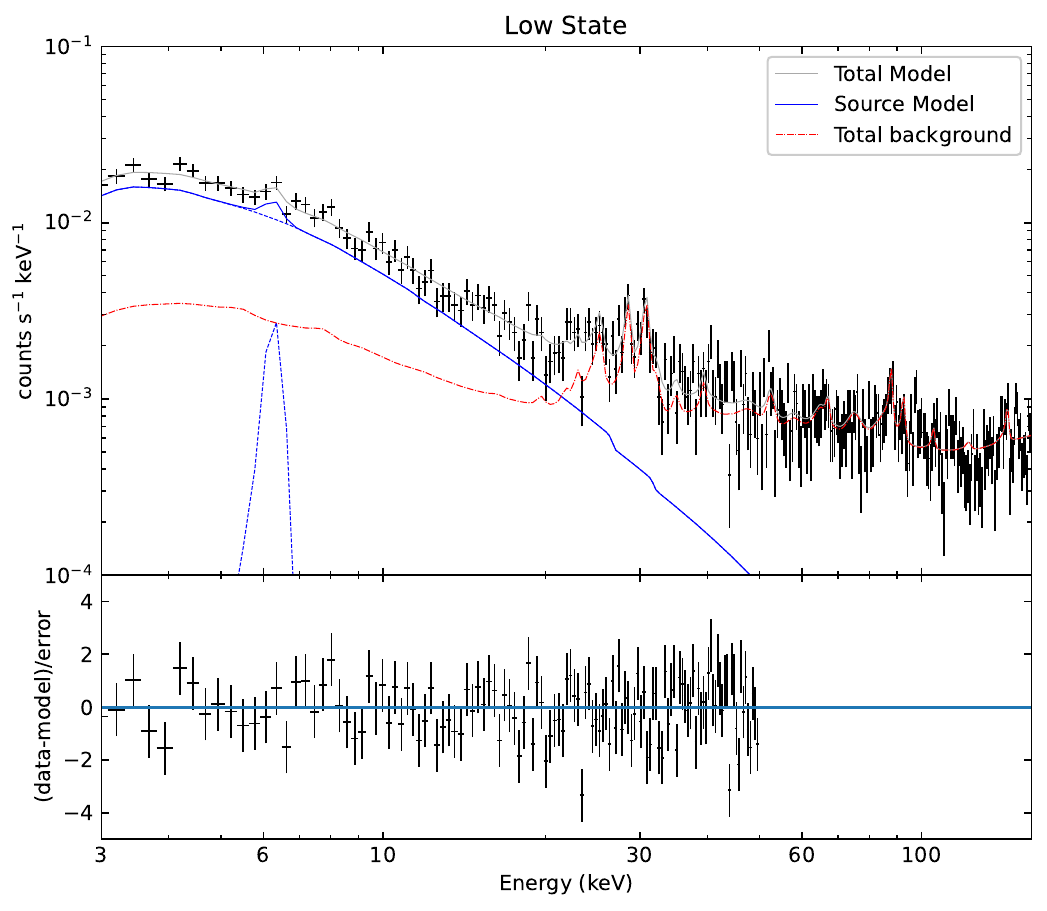}{0.46\textwidth}{}
    }
    \caption{Same as Figure \ref{fig:low-focused-spectra}, for stray light observation 30001130002. The solid blue line shows the total source model, with components as dashed lines; the solid red line indicates the background model from \texttt{nuskybgd}. The gray line is the sum of the source and background models. In the bottom panel of the low state spectrum, residuals are only shown for the energy range (3-20 keV) over which a fit of the source model was performed.}
    \label{fig:low-sl-spectra}
\end{figure*}

When not flaring, the 4U 1700-377's spectrum in stray light barely exceeds the background (see Fig.\ \ref{fig:low-sl-spectra}), leading to limited statistics for this state. In order to meaningfully compare this spectrum with the flaring state, it is necessary to freeze the parameters for the exponential cutoff portion of the power law to those found in the focused low state. The parameters for this model, and the model for the flaring state, are reported in Table \ref{tab:low-sl-models}. Accounting for uncertainties, photon index for the high- and low-flux states of the stray light observation shows a much smaller variation than that of the focused observation. At $90\%$ confidence, the photon index of the low-state observation may only be softer by 0.1. This low-state photon index is consistent with that found in the focused observation, as well. Interestingly the low-state spectrum fits with a much lower column density, consistent with just the galactic absorption of $\sim 0.5 \times 10^{22}$ cm$^{-2}$ \citep{HI4PI2016}.

In the stray light flaring spectrum, we check for any indications that a higher energy emission line or CRSF absorption feature might improve the \texttt{highecut} model. Searching between 3-10 keV, we do not find any indication that an emission line other than iron K$\alpha$ would improve the fit. Checking for an absorption feature, we find that a weak, narrow absorption feature can be fit around $\sim 13$ keV, however its strength is consistent with zero, as discussed in Section \ref{subsec:absorption-feature}. 

%% 30001130002A flare/low spectra

%% 30001130002A fit params for flare/low 
\begin{deluxetable}{lcc}[t!] \label{tab:low-sl-models}
\tablecaption{Flaring and persistent states of the stray light observation 30001130002A, modeled with \texttt{highecut}.}
\tablehead{\colhead{Param.} & \colhead{Flaring} & \colhead{Persistent}} 
\startdata
$N_H\ (10^{22}$ cm$^{-2}$) & 7.2$_{-2.4}^{+3.6}$    & $< 2.9$ \\
Photon Index               & 1.4$\pm0.1$            & $1.7\pm0.1$ \\
$E_{cutoff}$ (keV)         & $9\pm1$                & 7.9\tablenotemark{f} \\
$E_{fold}$ (keV)           & 30$_{-4}^{+5}$         & 31.0\tablenotemark{f} \\
kT (bbody, keV)            & $0.3_{-0.2}^{+0.1}$    & ... \\
$E_{Fe K\alpha}$ (keV)     & $6.4\pm0.1$            & $6.3\pm0.2$ \\
$\sigma_{Fe K\alpha}$ (eV) & 100\tablenotemark{f}   & 100\tablenotemark{f} \\ \hline
$F_{12}$\tablenotemark{g}  & $2.8^{+0.4}_{-0.2}$    & $0.46^{+0.04}_{-0.03}$ \\
$F_{50}$\tablenotemark{h}  & $6.6^{+0.4}_{-0.3}$    & $0.86 \pm 0.05$ \\ \hline
$C$ / d.o.f.               & 204.8362 / 187         & 137.0731 / 123 \\
\enddata
\tablenotetext{f}{Parameter has been frozen.}
\tablenotetext{g}{3-12 keV unabsorbed flux, $\times 10^{-9}$ ergs cm$^{-2}$ s$^{-1}$}
\tablenotetext{h}{3-50 keV unabsorbed flux, $\times 10^{-9}$ ergs cm$^{-2}$ s$^{-1}$}
\end{deluxetable}

%%%%%%%%%%%%%%%%%%%%%%%%%%%%%%%%%%%%%%%%%%%%%%%%%%%%%%%%%%%%%%%%%%%%%%%%%%%
%%%%%%%%%%%%%%%%%%%%%%%%%%%%%%%%%%%%%%%%%%%%%%%%%%%%%%%%%%%%%%%%%%%%%%%%%%%
\section{Discussion} \label{sec:discussion}

While two focused NuSTAR observations provide high quality spectral data for 4U 1700-377, a high degree of variability in the source's luminosity means that no singular observation is able to tell a complete story of accretion behavior in the system. These focused observations indicate variability in 4U 1700-377's continuum and emission lines, which we further investigate by leveraging an inventory of serendipitous stray light observations from NuSTAR. This allows us to extend the baseline of data for the source across a larger range of time, as well as capture additional instances of transient behavior. We detect frequent flaring behavior, as has been previously reported in this source, as well as two intervals of up to 60 ks in length during which the source shows no significant flaring activity. Studying these time intervals allows us to better investigate the variability of 4U 1700-377.

We test several continuum models on the focused observations, and find that a \texttt{highecut} power law model is able to provide the best spectral fits. Comptonization model \texttt{thcomp} also fits very well to the continuum, however with notable residuals slightly above 7 keV. When these residuals are addressed, however, the fit quality is comparable to the \texttt{highecut} model. 

In all of our observations we detect the presence of a ``soft excess" \citep{Hickox2004} below 5 keV. Using a blackbody to describe this component of the spectrum, we find our models fit with an energy in the range of $0.2-0.4$ keV and a total luminosity on the order of $10^{36}-10^{38}$ ergs s$^{-1}$. This blackbody spectrum corresponds to an emitting area of a few hundred kilometers in radius, suggesting the previously reported ``soft excess" is emission from photoionized stellar wind in the vicinity of the compact object.

Previous observations have found the line of sight absorption to the source to be variable, with values above the predicted galactic hydrogen column density $0.5 \times 10^{22}$ cm$^{-2}$ \citep{HI4PI2016}. \citet{Chicharro2018} found the value of $N_H$ to increase significantly during a flare, consistent with an increased density of the stellar wind. A similar variation in column density due to the wind from the companion star is also reported in \citep{GG2015}. Changes in the absorption column are also reported over orbital timescales as well. \citet{Xiao2024} finds the absorbing column density to increase after $\phi \sim 0.6$, consistent with the presence of an accretion wake. Another theorized cause for the change in absorption column density could involve an accretion stream which funnels gas from an accretion disk \citep{Haberl1989}. For the both the focused and stray light NuSTAR observations, we note variable line of sight absorption, with hydrogen column densities ranging from $\sim 4 \times 10^{22}$ cm$^{-2}$ to $\sim 15 \times 10^{22}$ cm$^{-2}$. Generally our time averaged observations taken later in 4U 1700-377's orbit ($\phi \gtrsim 0.6$) report a greater column density, in agreement with previous works. 

To better understand the nature of the variability in the source, we look specifically for transient behavior in both the iron emission lines and for detection of a possible CRSF. In particular, emission in the iron region may prove highly valuable in tracking the process of mass transfer via the stellar wind. In the two focused observations, we find that while emission from Fe K$\alpha$ is strong and persistent, the profile of the emission line and the possible excess emission above 6.4 keV shows the possibility of significant variability across different observations of the source. While any given model-spectrum combination indicates confident detection of excess fluorescent emission in the 6.4-7.5 keV range, the NuSTAR data lack sufficient spectral resolution to constrain a coherent picture of variability through this emission.

Additional fluorescence lines like Ni K$\alpha$ are seen in some other high-mass X-ray binaries. Vela X-1 shows orbital phase-dependent appearance of fluorescence lines including Ni K$\alpha$ \citep{Goldstein2004, MN2014, Amato2021}. In Vela X-1, obscuration of the neutron star and its continuum X-rays by an accretion wake around orbital phase $\phi \sim 0.5$ allows for stronger detection of fluorescence lines \citep{Goldstein2004, MN2014}, which are seen in scattered photons similar to the eclipse \citet{Goldstein2004}. Confident detection of Ni K$\alpha$ in observation 30101027002 of 4U 1700-377, taken around $\phi \sim 0.6$, when an accretion wake has been predicted to intersect our line-of-sight \citep{Jiaswal2015a, Xiao2024}, is thus consistent with the behavior seen in Vela X-1. Better measurements of a possible Ni K$\alpha$ detection at earlier orbital phases, such as might be achieved with XRISM \citep{Terada2021}, would allow for much better measurements describing large-scale structure in the stellar wind and accretion of 4U 1700-377.

Previous study of 4U 1700-377 has found no conclusive determination surrounding various claimed cyclotron resonant scattering features \citep{Reynolds1999, Jiaswal2015a, Seifina2016, Bala2020}. Recently, both \citet{Xiao2024} and \citet{WO2026} (using Insight-HMXT and NuSTAR respectively) have presented more complete pictures of the various configuration of absorption lines that might be fit to the continuum of 4U 1700-377, depending on the choice of continuum model. The addition of four NuSTAR stray light observations with usable data to beyond 20 keV allows us to further test some of the claimed CRSFs: in particular, the strong detection claimed using NuSTAR by \citet{Bala2020}. By performing fits with various configurations (strength, width, and centroid) of the $\sim 16$ keV line, we report that the stray light data is unable to constrain any claim about the (non-)detection of this CRSF. Given the extent to which 4U 1700-377's continuum can be well-described without invoking a cyclotron absorption feature, and the inability to consistently constrain the detection of a CRSF, we suggest that this is not currently the best physical explanation for the source's spectrum.

We also utilize two time intervals without significant flares to investigate the persistent emission of 4U 1700-377. In the observations we consider, two show examples of extended low phases 40-60 ks in length. During this low state, source luminosity decreases by roughly a factor of six, accompanied by significant softening of the spectrum. 

4U 1700-377's low-state emission has been studied previously, although over shorter time scales (\citealt[][10 ks]{Boroson2003}; \citealt[][9 ks]{Chicharro2018}; \citealt[][30 ks]{Jiaswal2015a}). \citet{Chicharro2018} uses a $\sim 9$ ks interval between flares to describe a radiative cooling phase, as opposed to a more efficient Compton cooling while the source flares, brought on by the accretion of a dense ``clump" of stellar wind. \citet{Jiaswal2015a} attributes their $\sim 30$ ks low state with the presence of an accretion wake. It is likely that the stellar wind plays a large role in the shorter term fluctuations in 4U 1700-377's brightness, as shown in \citet{Ducci2009} or \citet{Chicharro2018}, although it is difficult to say the extent to which it is a factor in longer term low phases, such as what we have observed. 

Comparing the \texttt{thcomp} fits to the focused observation during its high and low states, we find that the increase in electron temperature and photon index both align with a decrease in the optical depth of the Comptonizing medium \citep{Zdziarski2020}. For a wind-fed source like 4U 1700-377 \citep{Hainich2020}, this broad continuum variation would be consistent with ingestion of a colder, denser clump of wind during the flaring state and accretion from the hotter, rarefied inter-clump medium during the low state \citep{Oskinova2012}. The much lower line-of-sight absorption during the low state of stray light observation 30001130002 is also consistent with the suggestion of a large, under-dense region in the stellar wind \citep{MN2017}. 
 
Still, a precise model of the physical processes driving emission during flaring and low states is difficult to constrain based on continuum shape alone \citep{BeckerWolff2007}. Confirming that such an effect alone is responsible for the extended low states observed in 4U 1700-377 would be provide insight towards the study of wind fed accretion in HMXBs, whereby continuum measurements may be able to track how density variations in the stellar wind directly affect emission even very close to the compact object. Beyond this, the extended low states we observe may also be tied to changes in the accretion process seen in 4U 1700-377, such as the transition from radiative cooling to Compton cooling observed in \citet{Chicharro2018}. Given the length of time we observe without significant flaring, it could also be possible that a larger scale change in the accretion process is required for describing some of the variability seen in the spectrum of 4U 1700-377. Without confident measurements of 4U 1700-377's spin period \citep[e.g.,][]{Xiao2024} or magnetic field strength, however, it is impossible to determine whether or not extended low-luminosity phases may represent something akin to the onset of a propeller regime \citep{Tsygankov2016}. 

% %%%%%%%%%%%%%%%%%%%%%%%%%%%%%%%%%%%%%%%%%%%%%%%%%%%%%%%%%%%%%%%%%%%%%%%%%%%
% %%%%%%%%%%%%%%%%%%%%%%%%%%%%%%%%%%%%%%%%%%%%%%%%%%%%%%%%%%%%%%%%%%%%%%%%%%%
\section{Conclusions} \label{sec:conclusions}

Using two focused NuSTAR observations and five stray light observations, we are able to study the behavior of 4U 1700-377 during seven separate binary orbits between May 2015 and January 2022. These observations show a frequent, significant variation in brightness characteristic of this source, as well as two extended intervals of low flux outside of eclipse. By studying this longer baseline of observations we confirm variable line of sight absorption, while photon indices of time averaged observations generally remain consistent. We confirm that the soft excess previously found in this source is present in all of our observations, and is consistent with emission from a photoionized stellar wind at a distance of a few hundred km from the compact object.

We find that both focused observations indicate the presence of additional emission features beyond 6.4 keV (with confidences in excess of $3\sigma$), although the specific feature seen can be variable and model dependent. If observed using an instrument with sufficient spectral resolution, the possibility of detecting and constraining further fluorescent emission lines in the source outside of eclipse is likely to be a valuable tool for describing the stellar wind and large-scale accretion structure. 

The inclusion of the stray light data allows us to further test previously reported claims of CRSF detections and non-detection. In general, the addition of a cyclotron line in the spectrum of 4U 1700-377 does not significantly improve the fit of standard continuum models. This analysis of the possible CRSF in 4U 1700-377 as seen with stray light is consistent with the most recent analysis of NuSTAR's focused observations \citep{WO2026}. We do not claim the detection of a cyclotron line in the spectrum of 4U 1700-377.

From the set of observations, two instances of extended low-flux emission are characterized by a decrease in luminosity of around six times, and changes in the source's spectral shape. At up to 60 ks in length, these phases are longer than have been previously reported and may allow for a better understanding of the relationship between the infalling stellar wind and accretion processes near the compact object. 

In total, our analysis of these NuSTAR observations of 4U 1700-377 supports predictions about the variability in source brightness arising from a clumpy stellar wind, while also indicating possible longer term variability which warrants further study. We suggest that observations with a high spectral resolution instrument such as XRISM-Resolve would be highly valuable in measuring the variability of the stellar wind, its relationship with the accretion onto the compact object, and in constraining physical explanations for the variable emission line behavior we have reported.

\begin{acknowledgements}
    This research is makes use of data and tools hosted on the HEASARC online archive, provided by NASA/GSFC. This research has also made use of the NuSTAR Data Analysis Software (NuSTARDAS) jointly developed by the ASI Space Science Data Center (SSDC, Italy) and the California Institute of Technology (Caltech, USA). GM acknowledges support from the PRIN MUR SEAWIND (2022Y2T94C), funded by the European Union – Next Generation EU, Mission 4 Component 1 CUP C53D23001330006, and the INAF Grant BLOSSOM. We also thank the anonymous referee for their comments that have led to the improvement of this work.
\end{acknowledgements}

\bibliography{refs}{}
\bibliographystyle{aasjournal}

\appendix 
\restartappendixnumbering
\section{Products for the stray light observations} \label{appendix:allprods}

\begin{figure}[h!]
    \gridline{
    \fig{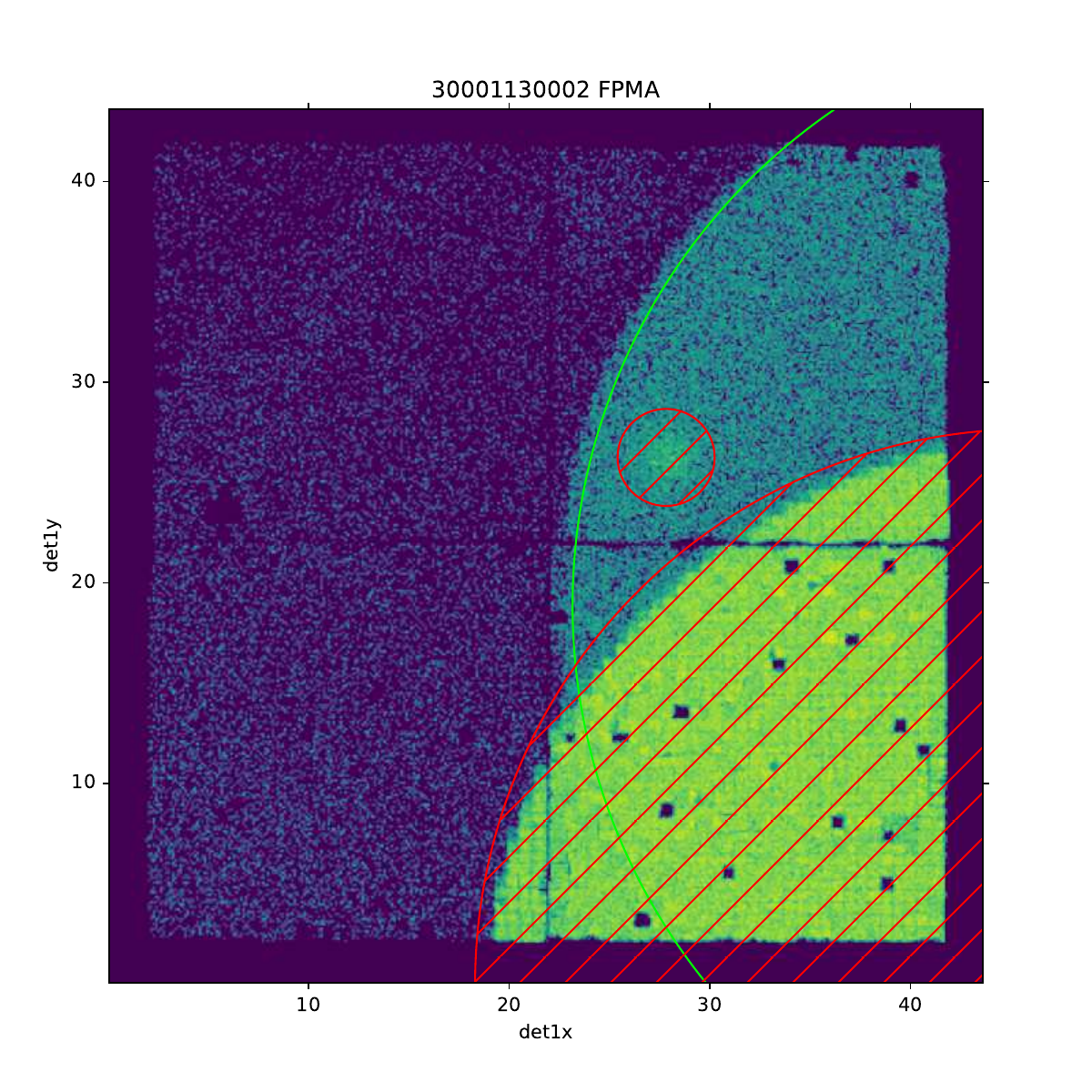}{0.4\textwidth}{}
    \fig{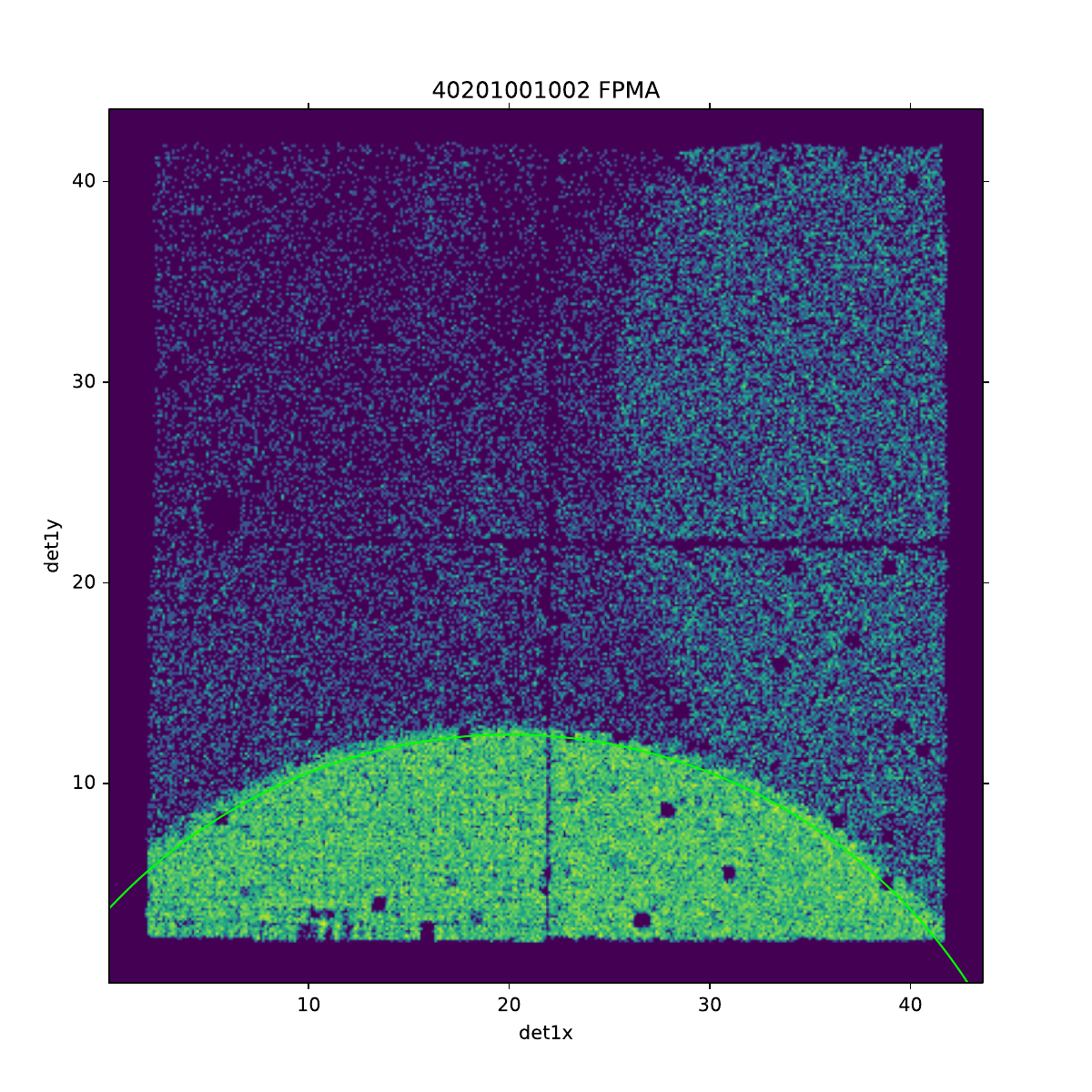}{0.4\textwidth}{}
    }
    \vspace{-1cm}
    \gridline{
    \fig{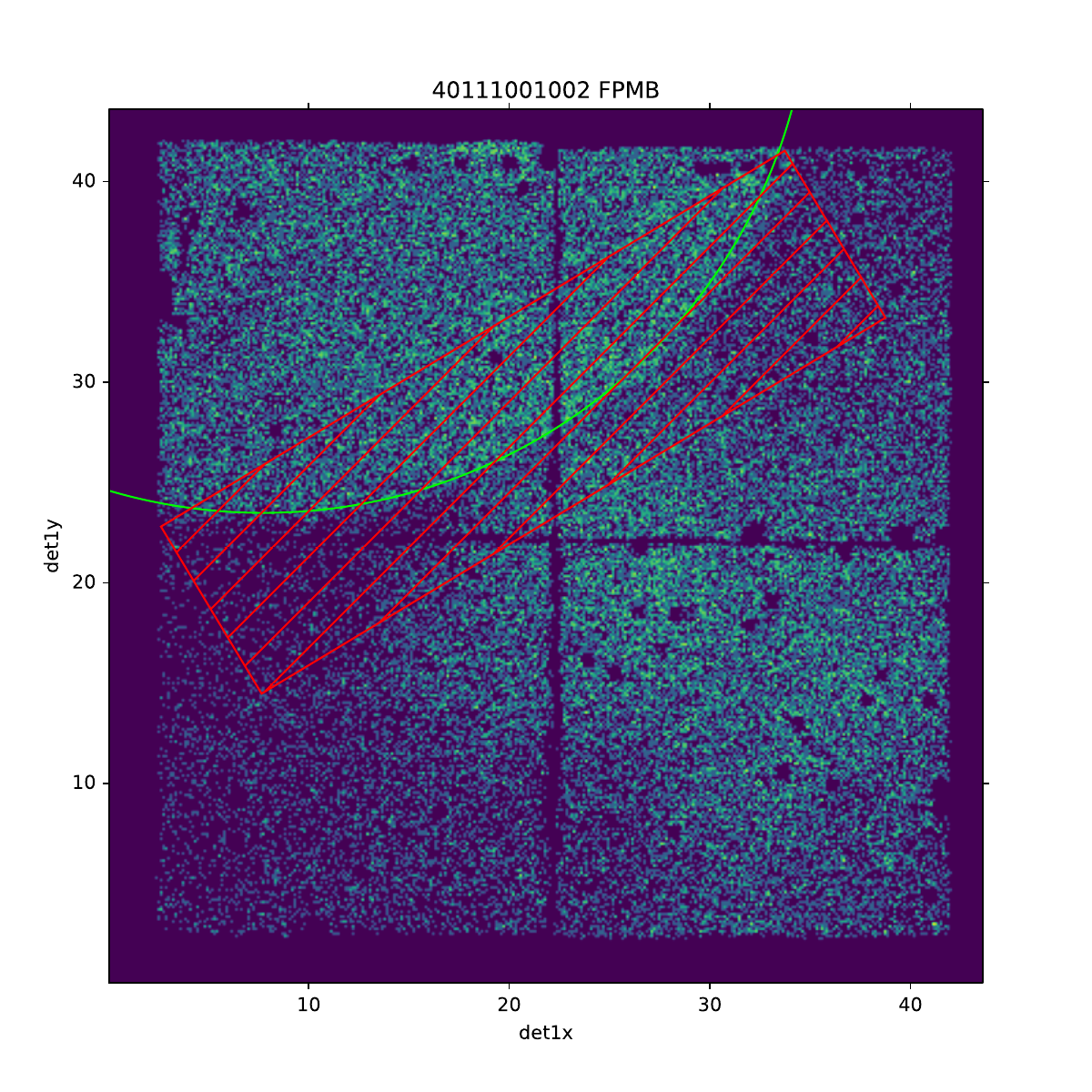}{0.4\textwidth}{}
    \fig{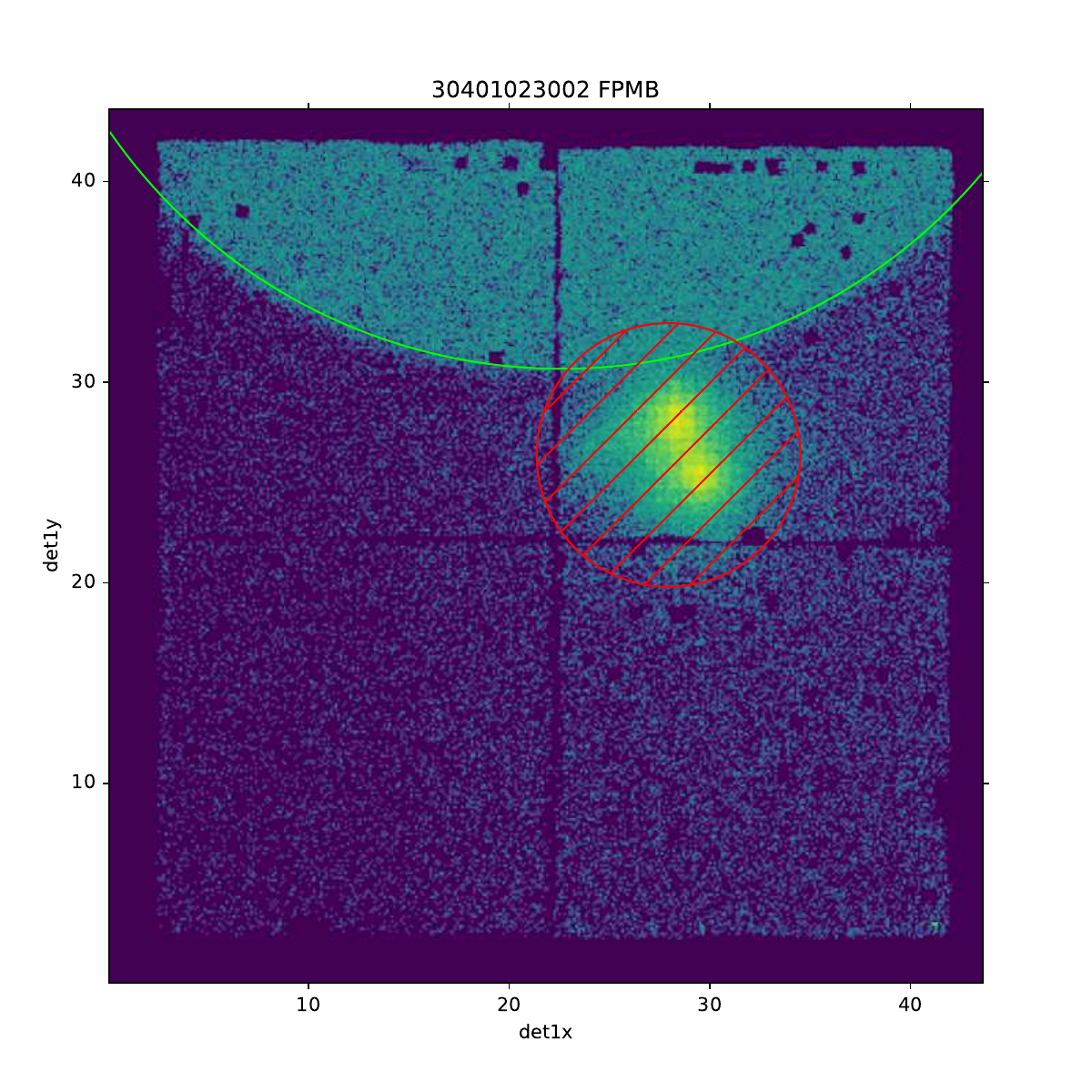}{0.4\textwidth}{}
    }
    \vspace{-1cm}
    \gridline{
    \fig{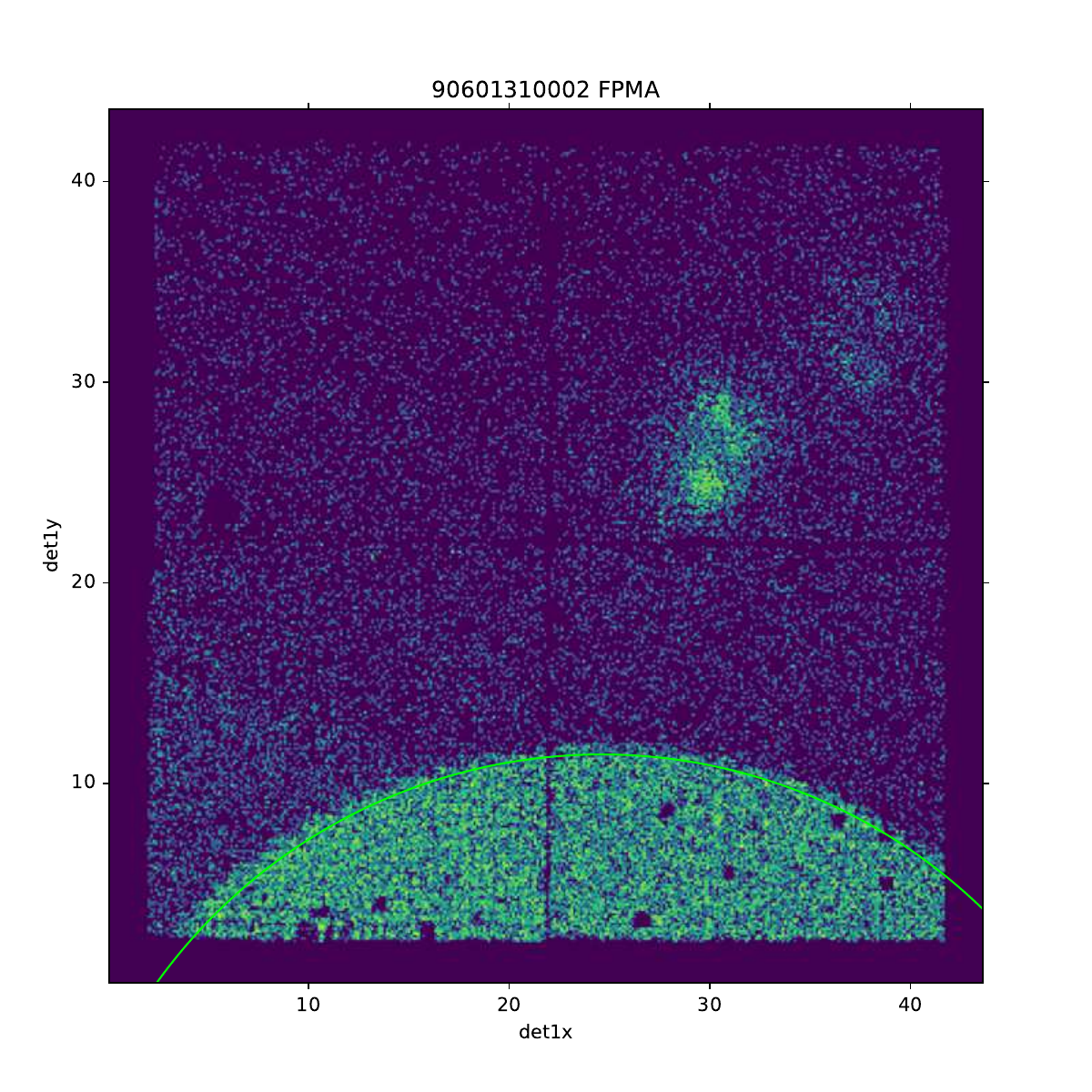}{0.4\textwidth}{}
    }
    \caption{Images for each analyzed stray light observation in detector coordinates. Extraction regions are identified in green; exclusion regions are shown in red. Color scaling is logarithmic. \label{fig:all-sl-images}}
\end{figure}

\begin{figure}[h!]
    \gridline{
    \fig{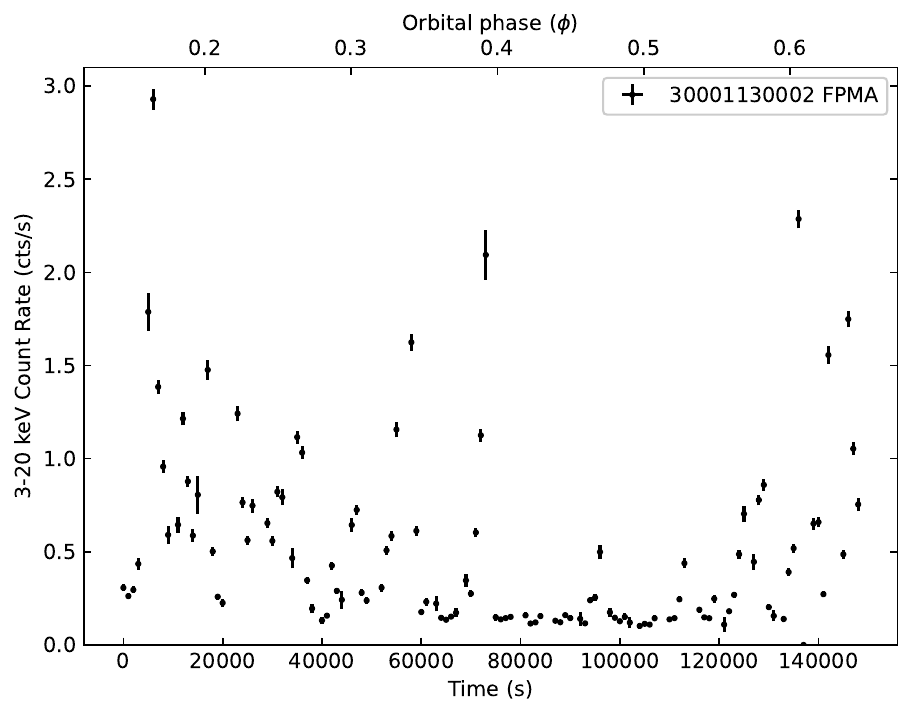}{0.45\textwidth}{}
    \fig{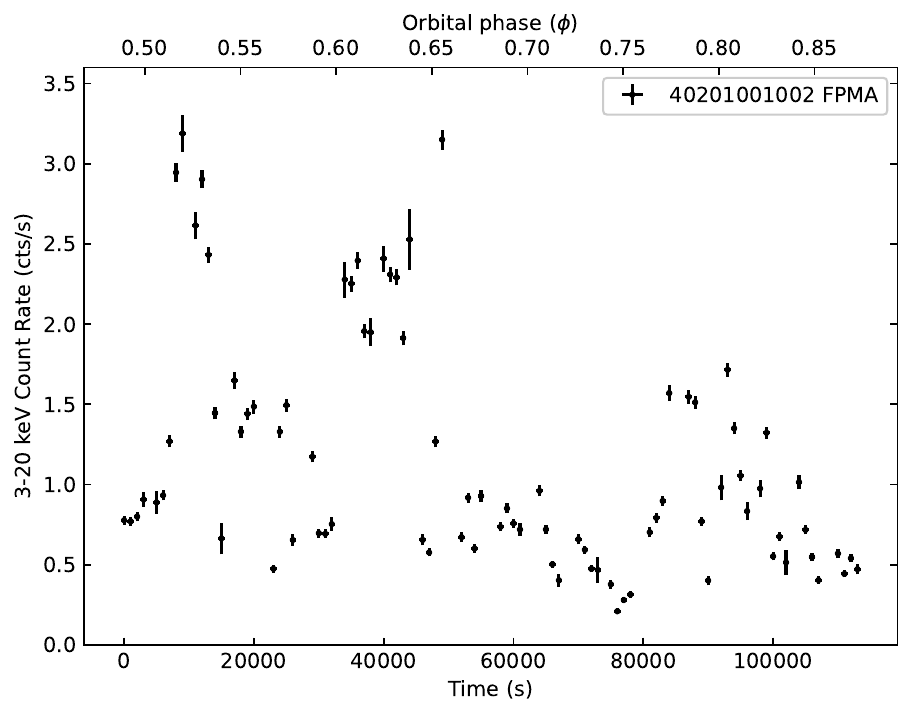}{0.45\textwidth}{}
    }
    \gridline{
    \fig{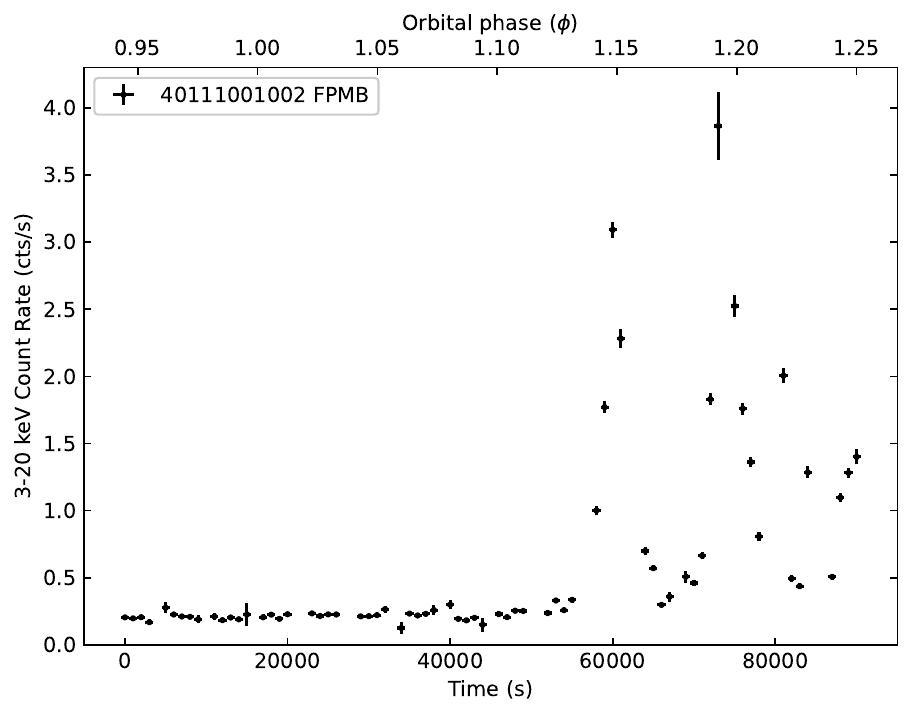}{0.45\textwidth}{}
    \fig{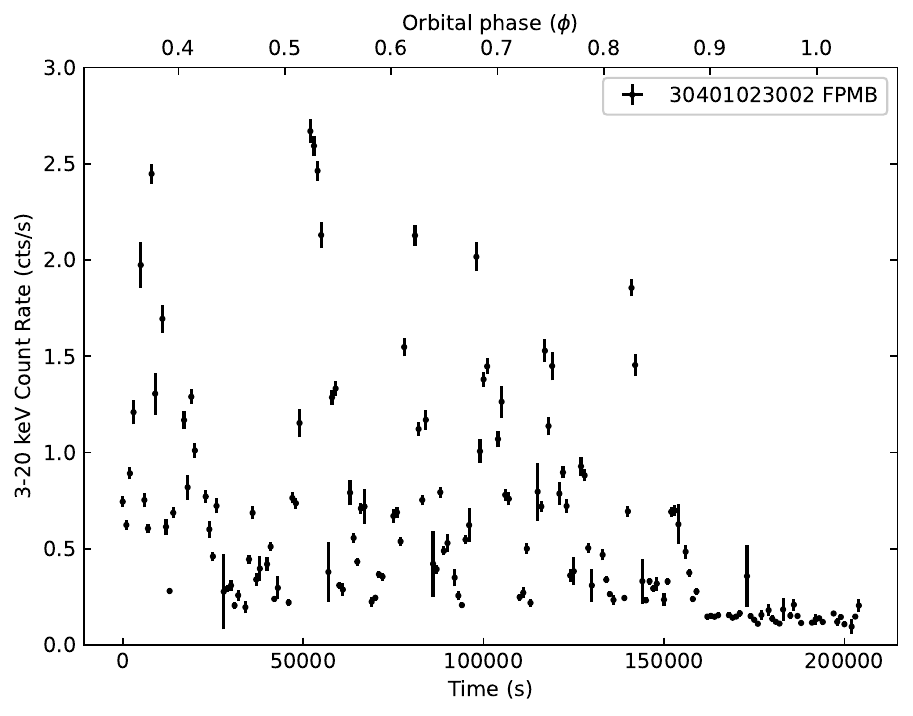}{0.45\textwidth}{}
    }
    \gridline{
    \fig{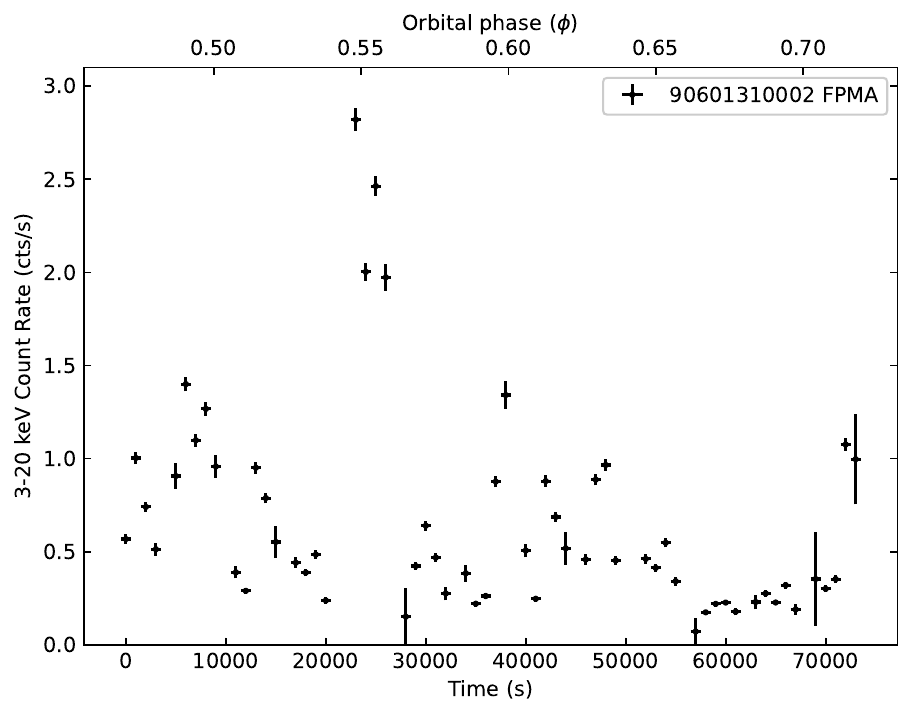}{0.45\textwidth}{}
    }
    \caption{Light curves for each analyzed stray light observation in energy range 3-20 keV, at 1000s bins. \label{fig:all-sl-lightcurves}}
\end{figure}

\begin{figure}[h!] 
    \gridline{
    \fig{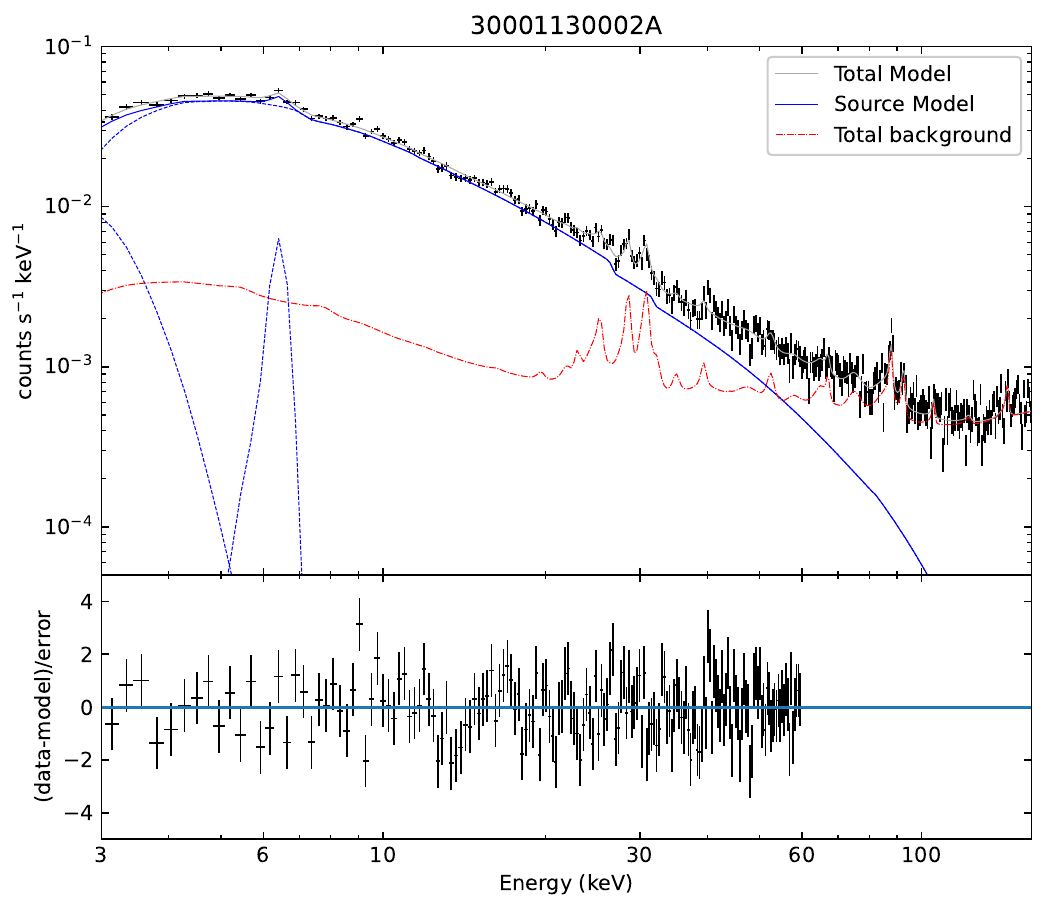}{0.4\textwidth}{}
    \fig{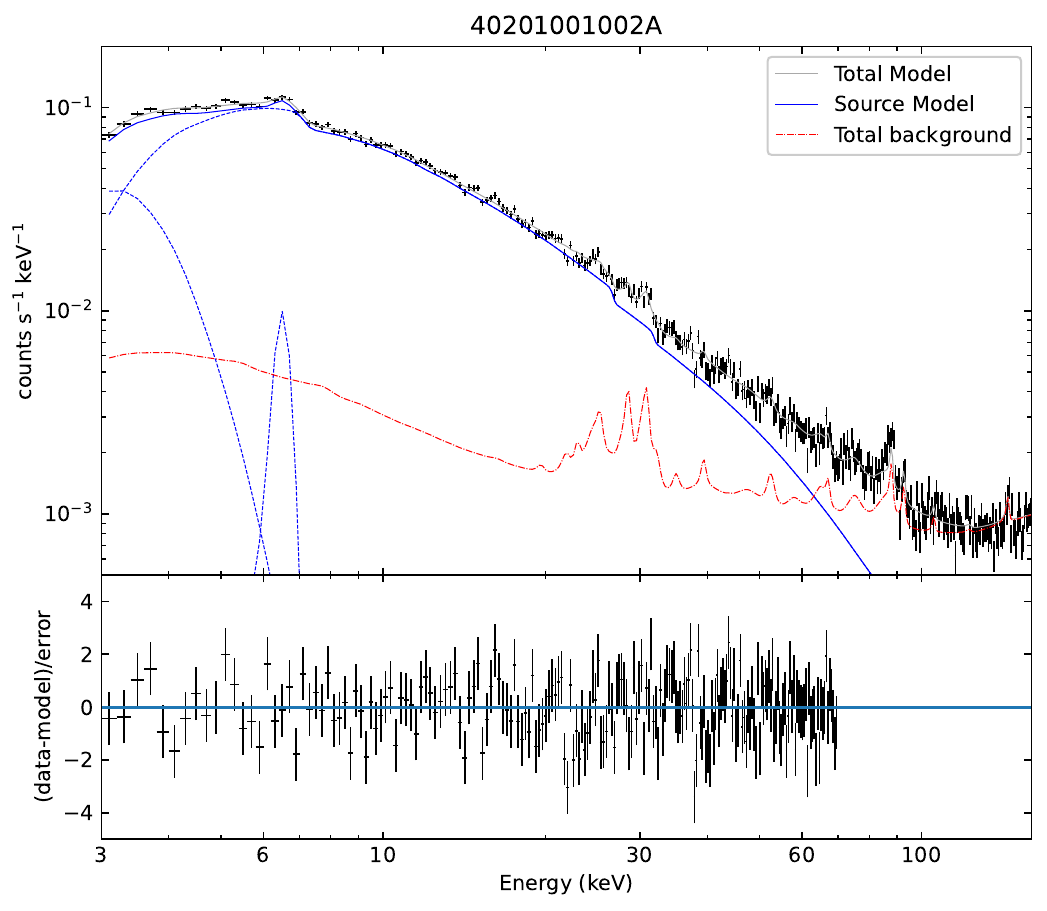}{0.4\textwidth}{}
    }
    \gridline{
    \fig{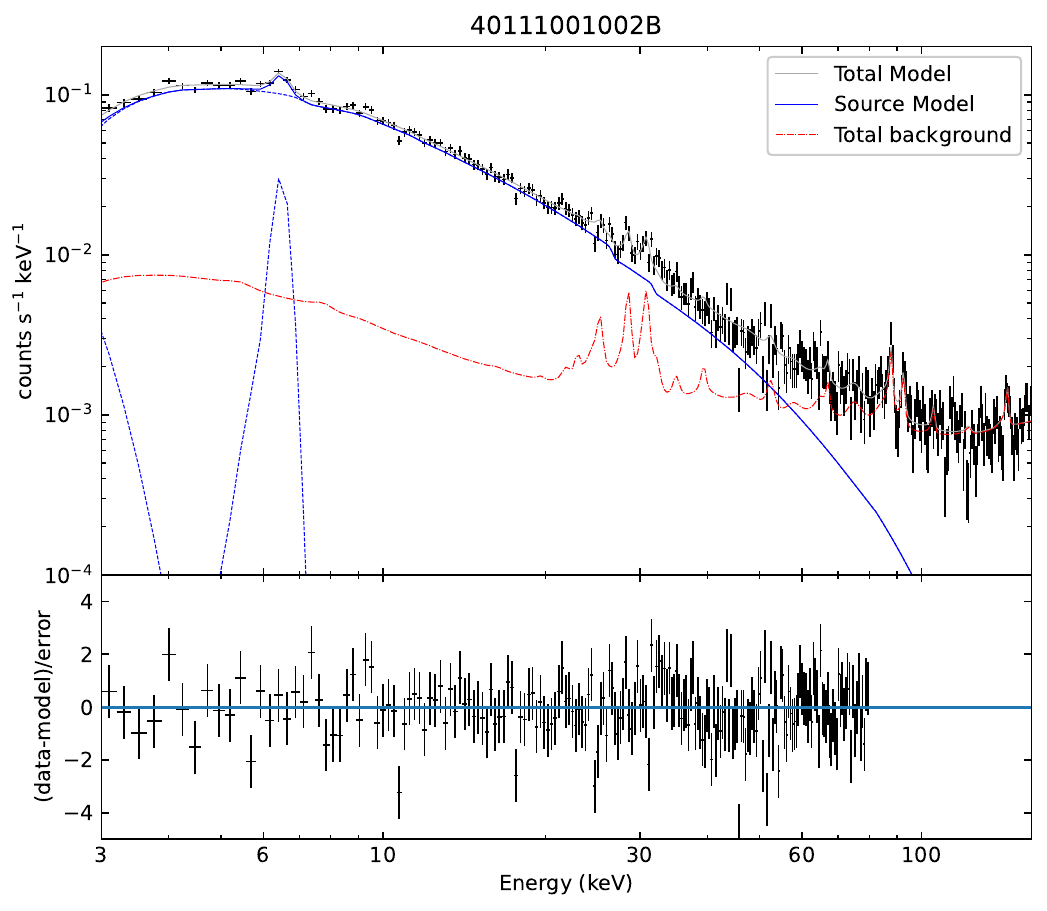}{0.4\textwidth}{}
    \fig{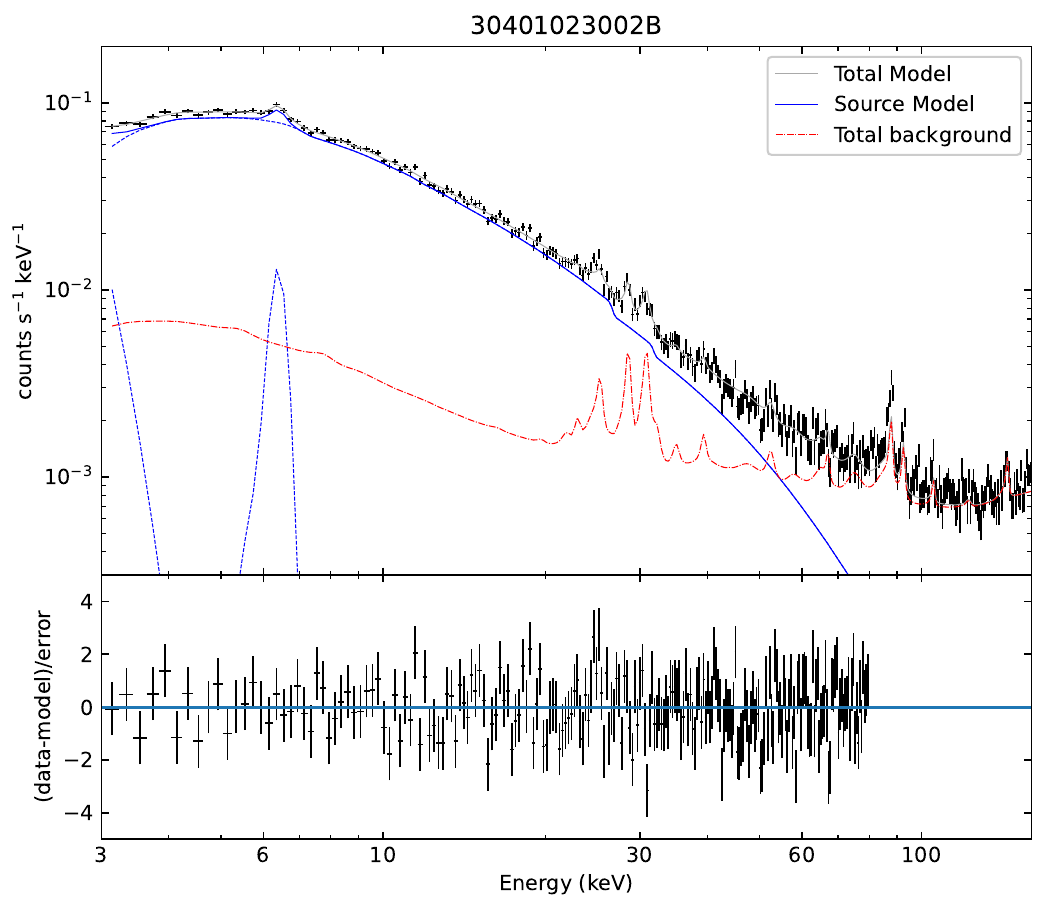}{0.4\textwidth}{}
    }
    \gridline{
    \fig{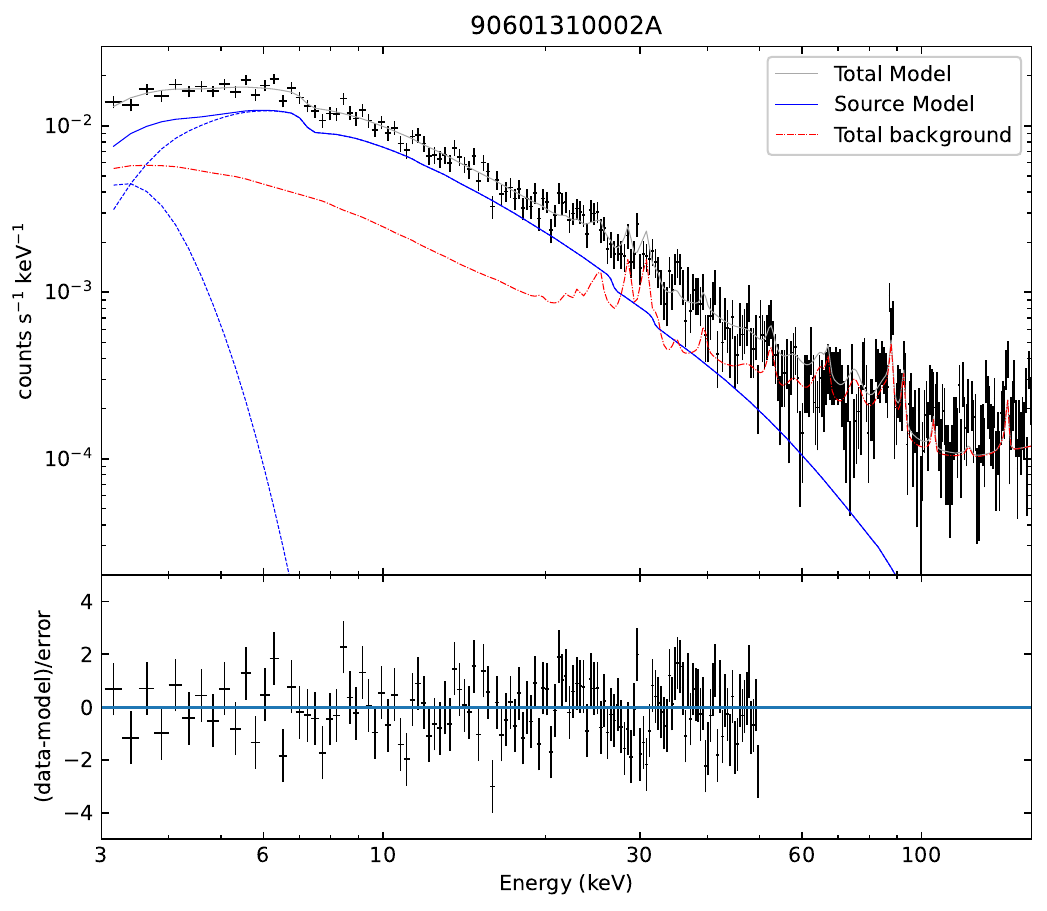}{0.4\textwidth}{}
    }
    \caption{Spectra and best fit model for all observation of 4U 1700-377 in the \texttt{StrayCats} catalog. A base model of \texttt{phabs(highecut*pow+bbody+gaus)} is applied to all spectra, as described in Section \ref{sec:spectral-analysis}. Components which do not improve the fit or whose parameters are unconstrained are removed from the model. The \texttt{nuskybgd} background is also shown as a dash-dot red line. The bottom panel shows the residuals of the best fit model computed as (data-model)/error. \label{fig:all-spectra}} 
\end{figure}

We show products from all of the analyzed stray light observations. Figure \ref{fig:all-sl-images} shows detector images from FPMA/B, along with selected stray light regions. Light curves for each observation are shown in Figure \ref{fig:all-sl-lightcurves}. Two observations (40111001002, 30401023002) include the eclipse of the compact object, which we have filtered out for our analysis. Figure \ref{fig:all-spectra} shows the spectrum for each observation, along with best fit residuals when modeled using a \texttt{highecut} continuum. The \texttt{nuskybgd} model \citep[][and described in Section \ref{sec:data-reduction}]{Wik2014} is shown in red, and agrees well with the source-region spectrum at high-energies where the background dominates. The parameters for these fits are reported in Table \ref{tab:all-sl-params}.

In general the fits to the time-averaged stray light data are consistent with the \texttt{highecut} models best fit to the focused observations. Most of the stray light observations report absorption column densities of $\sim 7 \times 10^{22}$ cm$^{-2}$, photon indices of $\sim1.3$, and a soft excess well-described by a blackbody of temperature $\sim 0.3$ keV. Two stray light observations (40201001002A and 90601310002A) reported absorption column densities $3-4$ times greater than the other observations, however. A prominent Fe K$\alpha$ fluorescence line is seen in four of the five stray light datasets. 

%% det1 images and regions for all SL obs

%% Light curves for all SL obs

%% Spectra from all stray light observations

%% Fit parameters for all stray light observations
\begin{deluxetable}{lccccc}[h!] \label{tab:all-sl-params}
\tablecaption{Fit parameters for the stray light observations modeled using \texttt{highecut}.}
\tablehead{\colhead{} & \colhead{30001130002A} & \colhead{40111001002B} & \colhead{40201001002A} & \colhead{30401023002B} & \colhead{90601310002A} } 
\startdata
$N_H\ (10^{22}$ cm$^{-2}$) & $7\pm 1$             & $5\pm1$              & $13_{-4}^{+3}$         & $4_{-1}^{+2}$          & $18\pm7$ \\
Photon Index               & 1.4$\pm0.1$          & 1.23$\pm 0.06$       & $1.3\pm0.1$            & $1.1\pm0.01$           & 1.5$\pm0.3$\\
$E_{cutoff}$ (keV)         & $9\pm1$              & $9\pm1$              & $7_{-2}^{+1}$          & $6.4\pm0.9$            & $9_{-6}^{+13}$ \\
$E_{fold}$ (keV)           & $34_{-5}^{+7}$       & 24$_{-2}^{+3}$       & $33_{-4}^{+5}$         & $22_{-2}^{+3}$         & $31_{-10}^{+20}$ \\
kT (bbody, keV)            & 0.3$_{-0.2}^{+0.1}$  & $< 0.17$             & $0.38_{-0.03}^{+0.04}$ & $0.2\pm0.1$            & $0.37_{-0.20}^{+0.02}$ \\
$E_{Fe K\alpha}$ (keV)     & 6.42$\pm0.09$        & 6.47$\pm 0.07$       & 6.5$\pm0.1$            & $6.39\pm0.07$          & ... \\
$\sigma_{Fe K\alpha}$ (eV) & 100\tablenotemark{f} & 100\tablenotemark{f} & 100\tablenotemark{f}   & 100\tablenotemark{f}   & ... \\ \hline
$F_{12}$\tablenotemark{g}  & $2.3^{+0.5}_{-0.2}$  & $3.0^{+0.4}_{-0.1}$  & $5.6^{+0.9}_{-0.7}$    & $2.7^{+0.3}_{-0.1}$    & $0.8^{+0.8}_{-0.3}$ \\
$F_{50}$\tablenotemark{h}  & $4.9^{+0.5}_{-0.3}$  & $7.4^{+0.4}_{-0.2}$  & $12.5^{+0.9}_{-0.7}$   & $6.8^{+0.3}_{-0.2}$    & $1.6^{+0.8}_{-0.3}$ \\ \hline
$C$ / d.o.f.               & 193.3565 / 168       & 219.1467 / 196       & 249.4306 / 200         & 255.7305 / 218         & 138.9286 / 129 \\
\enddata
\tablenotetext{f}{Parameter has been frozen.}
\tablenotetext{g}{3-12 keV unabsorbed flux, $\times 10^{-9}$ ergs cm$^{-2}$ s$^{-1}$}
\tablenotetext{h}{3-50 keV unabsorbed flux, $\times 10^{-9}$ ergs cm$^{-2}$ s$^{-1}$}
\end{deluxetable}

\end{document}